\documentclass[preprint]{aastex}
\usepackage{emulateapj5}
\usepackage{epsf}
\usepackage{psfig}
\usepackage{graphicx}
\usepackage{lscape}

\begin{document}
\title{Disk Loss and Disk Renewal Phases in Classical Be Stars I: Analysis of Long-Term Spectropolarimetric Data}
\author{John P. Wisniewski\altaffilmark{1,2}, Zachary H. Draper\altaffilmark{1}, 
Karen S. Bjorkman\altaffilmark{3}, Marilyn R. Meade\altaffilmark{4}, Jon E. Bjorkman\altaffilmark{3}, Adam F. Kowalski\altaffilmark{1}}

\altaffiltext{1}{Department of Astronomy, University of Washington, Box 351580 Seattle, WA 98195, USA, jwisniei@u.washington.edu, zhd@uw.edu, adamfk@u.washington.edu}
\altaffiltext{2}{NSF Astronomy \& Astrophysics Postdoctoral Fellow}  
\altaffiltext{3}{Ritter Observatory, Department of Physics \& Astronomy, Mail Stop 113, University of Toledo, Toledo, OH 43606, karen.bjorkman@utoledo.edu, jon@physics.utoledo.edu}
\altaffiltext{4}{Space Astronomy Lab, University of Wisconsin-Madison, 1150 University Avenue, Madison, WI 53706, meade@astro.wisc.edu}

\begin{abstract} Classical Be stars are known to occasionally transition from having a gaseous circumstellar disk (``Be phase'') to a state 
in which all observational evidence for the presence of these disks disappears (``normal B-star phase'').  We present one of the most comprehensive 
spectropolarimetric views to date of such a transition for two Be stars, $\pi$ Aquarii and 60 Cygni.  60 Cyg's disk loss episode 
was characterized by a generally monotonic decrease in emission strength over a time-scale of $\sim$1000 days from the maximum V-band polarization to the 
minimum H$\alpha$ equivalent width, consistent with the viscous time-scale of the disk, assuming $\alpha$ $\sim$0.14.  $\pi$ Aqr's disk 
loss was episodic in nature and occurred over a time-scale of $\sim$2440 days.  An observed time lag between the behavior of the polarization and H$\alpha$ in both stars indicates the disk clearing proceeded in an 
``inside-out'' manner.  We determine the position angle of the intrinsic polarization to be 166.7$\pm$0.1$^{\circ}$ for $\pi$ Aqr and 107.7$\pm$0.4$^{\circ}$ for 60 Cyg, and model the wavelength dependence of the observed polarization during the quiescent diskless phase of each star to determine the interstellar polarization along the line of sight.  Minor outbursts observed during the quiescent phase of each star 
shared similar lifetimes as those previously reported for $\mu$ Cen, suggesting that the outbursts represent the 
injection and subsequent viscous dissipation of individual blobs of material into the inner circumstellar environments of these stars.  We also observe deviations from the mean intrinsic polarization position angle during polarization outbursts in each star, indicating deviations from axisymmetry.  We propose that these deviations might be indicative of the injection (and subsequent circularization) of new blobs into the inner disk, either in the plane of the bulk of the disk material or in a slightly inclined (non-coplanar) orbit.

\end{abstract}


\section{Introduction} \label{intro}

Since the pioneering work of \citet{str31}, a sub-set of B-type main sequence stars rapidly rotating at velocities 
ranging from $>$40\% (for early-type B stars; \citealt{cra05}) to possibly 100\% of the critical rate \citep{tow04} have been
identified.  Photometric, spectroscopic, polarimetric, and interferometric observations of these so called classical Be stars indicate that they are characterized by the presence of a geometrically flattened, 
circumstellar decretion disk (for a recent review see \citealt{por03}).  While for single stars the origin of the gas in these disks is agreed to be material ejected 
from the stellar photosphere, the precise mechanism for forming these disks remains elusive.  Suggested disk formation mechanisms include 
the compression of stellar winds (``WCD'' model; \citealt{bjo93}), potentially aided by the channeling of magnetic fields \citep{cas02,bro08}, 
as well as the viscous decretion disk model \citep{lee91}, which could be fed via non-radial pulsational events \citep{cra09} similar to those 
which have been observed in several Be stars \citep{riv98,nei02}.

Classical Be stars are known to exhibit variability on a multitude of time-scales and amplitudes, including short-term (hours to weeks) 
polarimetric variability \citep{car07}, multi-periodic short-term spectroscopic variability arising from non-radial pulsations \citep{riv98}, 
and long-term (several to tens of years) cyclical spectroscopic variability originating from one-armed density waves in the disk 
\citep{oka91,ste09}.  Arguably the most dramatic type of variability observed is the aperiodic ``Be to normal B to Be'' transition whereby 
classical Be stars dissipate (i.e. lose) their disks, and will often begin to regenerate a new disk after a period of relative quiescence 
(see e.g. \citealt{und82} and references therein).  While several individual stars have been observed to experience this transition one or 
more times \citep{und82,doa83,cla03,vin06}, it is not known whether there is a characteristic time-scale for disk-loss and regeneration that would suggest a singular disk formation mechanism or 
whether each event is unique.  Although their sample size is small, recent monitoring of eight open clusters by \citet{mcs09} has 
revealed a surprising 12 new transient Be stars out of a sample of 296 stars monitored, suggesting that disk-loss/renewal episodes may be a prevalent phenomenon.  Be stars experiencing a ``Be to normal B to Be" transition are clearly of significant astrophysical importance as they provide observational evidence of how Be disks form, thereby constraining theoretical efforts to identify the disk formation mechanism.

Polarimetry has an established history of providing a unique diagnostic in identifying and studying the detailed circumstellar environments 
of both individual classical Be stars \citep{qui97,woo97,cla98} and larger statistical ensembles \citep{coy69,mcl78,poe79,wi07b}.  Electron 
scattering modified by pre- or post-scattering absorption by hydrogen atoms in the disk produces the characteristic wavelength dependent 
polarization signature of Be stars \citep{wo96a,wo96b}.  Modeling such data yields information about the density in the inner region of these 
disks \citep{woo97,car09} and has been used, along with interferometric observations, to confirm that the disks are geometrically thin 
\citep{qui97}. 

The two stars we examine in this paper, 60 Cygni and $\pi$ Aquarii, are well known 
classical Be stars.  60 Cyg (HD 200310) is a B1Ve classical Be star \citep{sle82}.  Periodic radial velocity variations of H$\alpha$ and He I 6678 \AA\ suggest that the star might be a spectroscopic binary having a period of 146.6 $\pm$0.6 days, while 1.064 day line profile 
variations and 0.2997 day photometric variations may be caused by non-radial pulsations \citep{kou00}.  \citet{kou00} present evidence 
that the star has experienced disk-loss and subsequent regeneration in the past (see e.g. their Figure 1), with photometric variability 
positively correlated with H$\alpha$ emission strength; however, they neither discuss the detailed time-scale for these events nor 
publish their data in sufficient format to derive this information.  $\pi$ Aqr (HD 212571) is a B1 III-IVe classical Be star \citep{sle82}.  
As discussed in \citet{bjo02} and references therein, $\pi$ Aqr's most recent disk likely developed in the early 1950s and persisted until 
1996.  Analysis of spectroscopic data obtained after 1996, when the star was in a diskless ``normal B star'' phase, revealed the star has 
a close binary companion with an orbital period of 84.1 days \citep{bjo02}, which might be engaged in variable mass transfer with the primary.

In this paper, we present and analyze an extensive spectropolarimetric dataset which covers the 
disk-loss phase transition in two Be stars.  In Section \ref{data}, we outline our new spectroscopic and spectropolarimetric observations of 
60 Cyg and $\pi$ Aqr.  We analyze the time evolution of our spectroscopic data in Section \ref{halpha} and the time evolution of our polarimetric 
data in Section \ref{pol}.  A thorough discussion of the interstellar polarization along the line of 
sight to 60 Cyg and $\pi$ Aqr is presented in Section \ref{isp}.  Finally we offer a detailed discussion and interpretation of our results in Section \ref{discussion} and suggest future work in Section \ref{future}.

\section{Observations and Data Reduction} \label{data}

The spectropolarimetric data analyzed in this study were obtained by the University of
Wisconsin's (UW) HPOL spectropolarimeter, mounted on UW's 0.9m 
Pine Bluff Observatory (PBO) telescope.  $\pi$ Aquarii was observed on 127 nights between 1989 August 8 and 2004 
October 10 while 60 Cygni was observed on 35 nights between 1992 August 3 and 2004 September 26 (see Tables \ref{60cygdata} and 
\ref{piaqrdata}).  
Data obtained before 1995 were recorded using a dual Reticon array detector spanning the wavelength
range of 3200 -7600\AA\ with a spectral resolution of 25\AA\    
 \citep{wol96}.  Since 1995, HPOL has been equipped with a 400 x 1200 pixel CCD 
 and two gratings which provide coverage from 3200 -6020\AA\ and 5980 - 10,500\AA\ 
 at a resolution of 7\AA\ below 6000\AA\ and 10\AA\ above this point \citep{nor96}.
Further details about HPOL can be found in \citet{noo90}, \citet{wol96}, and \citet{har00}, and all HPOL data obtained between 1989-1994 are publically 
available from the Multimission Archive at STScI (MAST) website: http://archive.stsci.edu/hpol.

HPOL data were reduced, processed, and analyzed using REDUCE, a spectropolarimetric software 
package developed by the University of Wisconsin-Madison (see\citealt{wol96}).  Routine monitoring of unpolarized standard stars at PBO 
has enabled the instrumental polarization to be carefully calibrated; the residual instrumental systematic errors depend mildy on the date of 
the observations, but range from 0.027-0.095\% in the U-band, 0.005-0.020\% in the V-band, and 0.007-0.022\% in the I-band.
HPOL spectroscopic data are not calibrated to an absolute flux level due to the nonphotometric skies typically present during the 
observations \citep{har00}.

We supplement our HPOL spectroscopic observations of 60 Cyg with 65 nights of spectra obtained from 1998 August 26 to 2006 December 9 
using the fiber-fed echelle spectrograph of the 1.0m Ritter Observatory at the University of Toledo (Table \ref{60cygdata}).  These spectra span nine non-adjacent 70 \AA\ wide orders in the range 5285-6595\AA\, with R $\sim$26,000.  The data were reduced in IRAF using standard spectroscopic techniques, including bias correction, flat-fielding, and wavelength calibration.  Additional details about Ritter data are given in \citet{mor97}.  

Our HPOL polarimetric observations of $\pi$ Aqr are also supplemented by literature V-band polarization measurements (Table \ref{piaqrew}; \citealt{mcd86,mcd90,mcd94,mcd99}).  Moreover, both our 60 Cyg and $\pi$ Aqr data are supplemented by unpublished observations 
made with the Lyot Polarimeter at PBO betwen 1979-1987 (Table \ref{lyot}), which spanned an effective wavelength range of 4800-7600 \AA\ in 1979, 
4800-8200 \AA\ in 1980, and 4600-7200 \AA\ in 1987.  Further details about the PBO Lyot Polarimeter can be found in \citet{lup87, whi89}; and \citet{no90b}.

\section{Data Analysis}

\subsection{H-$\alpha$ Spectroscopy} \label{halpha}

We begin our analysis by determining the H$\alpha$ equivalent width (EW) of all of our spectroscopic data, with no correction for 
the constant underlying photospheric absorption line present.  Following continuum normalization, the EWs of our Ritter data were 
computed using \textit{splot} in IRAF\footnote{IRAF is distributed by the National Optical Astronomy Observatory, which is operated by the Association of Universities for Research in Astronomy, Inc., under contract with the National Science Foundation.}.  Equivalent width errors 
were computed using $\sigma^{2}$ = N[h$_{\lambda}$/(S/N)]$^{2}$(f$_{*}$/f$_{c}$), where N is the number of pixels across the H$\alpha$ line, h$_{\lambda}$ is the dispersion in \AA$^{-1}$, f$_{*}$ is the flux in the line, f$_{c}$ is the flux in the continuum, and S/N is the signal-to-noise ratio calculated in a line-free region of the spectra neighboring the H$\alpha$ line \citep{cha83}.  For our HPOL data, the spectral 
regions 6350-6450 and 6700-6800\AA\ were used to constrain the continuum flux around H$\alpha$ while the region 6520-6620\AA\ was used to measure the H$\alpha$ line flux.  Three nights of our $\pi$ Aqr HPOL dataset (MJD 2452860, 2452906, 2453181) suffered from 
extremely poor wavelength calibration, requiring us to shift the three spectra from these nights by $\sim$80\AA.  

Our EW data are compiled in Table \ref{60cygdata} (60 Cyg) and Table \ref{piaqrdata} ($\pi$ Aqr).  The EWs of our HPOL and Ritter observations 
of 60 Cyg are systematically offset from one another.  This is likely caused both by the different resolution of the instruments used and by the continuum in our low-resolution HPOL spectra not being perfectly fit via the linear function derived from our 2 continuum filters.  We stress however that this constant offset has no effect on our analysis or conclusions.  Comparison of four nights in which both instruments obtained spectroscopic observations indicates this offset is $\sim$1.9\AA; for plotting purposes only (Figure \ref{60cygew}), we have added a constant offset of 1.9\AA\ to our HPOL EWs measurement.  Note that no offset has been added to the tabulated measurements in Table \ref{60cygdata}.

\subsubsection{60 Cyg} \label{60cyghalpha}

Figure \ref{60cygew} clearly demonstrates that 60 Cyg experienced a fundamental change at H$\alpha$ throughout our 14 years of spectroscopic monitoring data, during which the line changed from absorption to emission to absorption.  \citet{kou00} qualitatively report 
a potentially similar event prior to our observations, but they did not publish their detailed EW measurements and only referred to the disk 
as being in ``strong emission, an intermediate state, and pure absorption'', thereby precluding a more comprehensive comparison of the 
events.  Our data indicate that it took $\sim$870 days for 60 Cyg's disk to transform from its strongest H$\alpha$ emission state (1998 December 24; Table \ref{60cygdata}) to its lowest state of pure absorption, the latter of which we interpret as the time when the disk had completely 
dissipated (2001 May 9; Table \ref{60cygdata}).  Inspection of the decline state, seen in the right-panel of Figure \ref{60cygew}, reveals that to first order the disk's emission strength followed a monotonic decrease void of any dramatic outbursts.  After reaching its low state in 2001 May, 60 Cyg remained at a moderately low state characterized by occasional 
minor increases in the H$\alpha$ EW through the last of our observations in 2006 December.

Binary companions can influence the circumstellar disks of Be stars in several ways.  They are believed to influence the behavior of global disk oscillations \citep{okt09}, truncate the outer disk \citep{car09}, and potentially trigger mass-loss events from the central star for systems in very eccentric orbits (e.g. $\delta$ Sco; \citealt{mir01}).  Assuming the range of primary and secondary masses for 60 Cyg given by \citet{kou00} and a circular orbit, 60 Cyg's disk could be truncated at the tidal distance of $\sim$120-135 R$_{sun}$, assuming the tidal radius is $\sim$0.9 R$_{Roche}$ (valid for small mass ratios; \citealt{whi91}) and using the Roche radius approximation of \citet{egg83}.  Aside from this possible truncation, we suggest it is unlikely that the binary companion is a fundamental contributor to 60 Cyg's disk-loss event.  The time-scale of the disk-loss event ($\sim$870 days) corresponds to almost 6 complete orbits of the binary companion (assuming a 146.6 day period; \citealt{kou00}), suggesting that the binary does not influence the primary star (or its disk) in a manner similar to the highly eccentric $\delta$ Sco system \citep{mir01}.

\subsubsection{$\pi$ Aqr} \label{piaqrhalpha}

The time evolution of the H$\alpha$ EW of $\pi$ Aqr (Figure \ref{piaqrew}) also provides clear evidence of a disk-loss 
event, at a higher sampling frequency than available in previous publications \citep{bjo02}.  Our data indicate that it took at least $\sim$2440 days 
for $\pi$ Aqr's disk to transform from its strongest state observed during our time coverage (1989 October 26; EW = -24.6 \AA\ Table \ref{piaqrdata}) to the likely start of 
its minimum state of pure absorption (1996 July 4; EW = 4.1 \AA\ Table \ref{piaqrdata}).  While the overall behavior of $\pi$ Aqr's decline phase shows a mostly 
monotonic decline in emission strength (left panel, Figure \ref{piaqrew}), similar to that seen for 60 Cyg, there is clear evidence of two 
significant interruptions to this trend.  The decline of $\pi$ Aqr's H$\alpha$ emission experienced a $\sim$220 day flattening event beginning on 1991 May 12 (JD 2448389) and a $\sim$200 day event beginning on 1994 June 5 (JD 2449509), during which its EW not only 
flattened but exhibited evidence of temporarily increasing in emission strength (right panel, Figure \ref{piaqrew}).

The time-scale for $\pi$ Aqr's disk-loss episode ($\sim$2440 days) corresponds to roughly 29 complete orbits of its known binary companion (assuming a 84.1 day period; \citealt{bjo02}), which suggests that the passage of the binary was not responsible for triggering the disk-loss episode.  Assuming the range of primary and secondary masses for $\pi$ Aqr given by \citet{bjo02} and a circular orbit, we calculate that $\pi$ Aqr's disk could be truncated at the tidal distance of 80-89 R$_{sun}$, assuming the tidal radius is $\sim$0.9 R$_{Roche}$ \citep{whi91} and using the Roche radius 
approximation of \citet{egg83}.  Aside from this possible truncation, we suggest it is unlikely that the binary companion influences the primary star (or its disk) in a manner similar to the highly eccentric $\delta$ Sco system \citep{mir01}.

After the likely start of its minimum ``pure absorption'' state in 1996 July, $\pi$ Aqr's H$\alpha$ EW remained generally unchanged until 2001 November, at which time a minor filling-in of the photospheric absorption line occurred and lasted until 2002 June.  A similar strengthening 
event appeared to begin at or around 2003 October and persisted through our final observation on 2004 October, though the H$\alpha$ 
emission strength never approached the values observed during the star's previous strong-disk phase.

\subsection{Total V-band Polarization} \label{pol}

Theoretical models predict that the predominant physical locations in the disk at which H$\alpha$ photons are emitted and at which photons are scattered to produce 
the observed polarization differs in Be stars, with the polarization produced in the very inner disk region and H$\alpha$ photons 
coming from throughout a much more extended radial distance \citep{car06,car09}.  For example, application of these models to observational data 
for the Be star $\zeta$ Tau \citep{wi07a,ste09} suggest that while the scattering events producing polarized photons are produced within 
a few stellar radii, H$\alpha$ photons are emitted over a substantially broader region of the disk, extending out to $\sim$40 stellar radii.  
Therefore, our spectropolarimetric dataset provides unique simultaneous insight into the long-term behavior of the inner versus outer regions of 60 Cyg's and $\pi$ Aqr's circumstellar disks.  To facilitate comparison with our H$\alpha$ EW data, we have binned our spectropolarimetric data to replicate the Johnson V-band filter.

\subsubsection{60 Cyg} \label{60cygpol}

\begin{figure}
\begin{center}
\includegraphics[width=8cm]{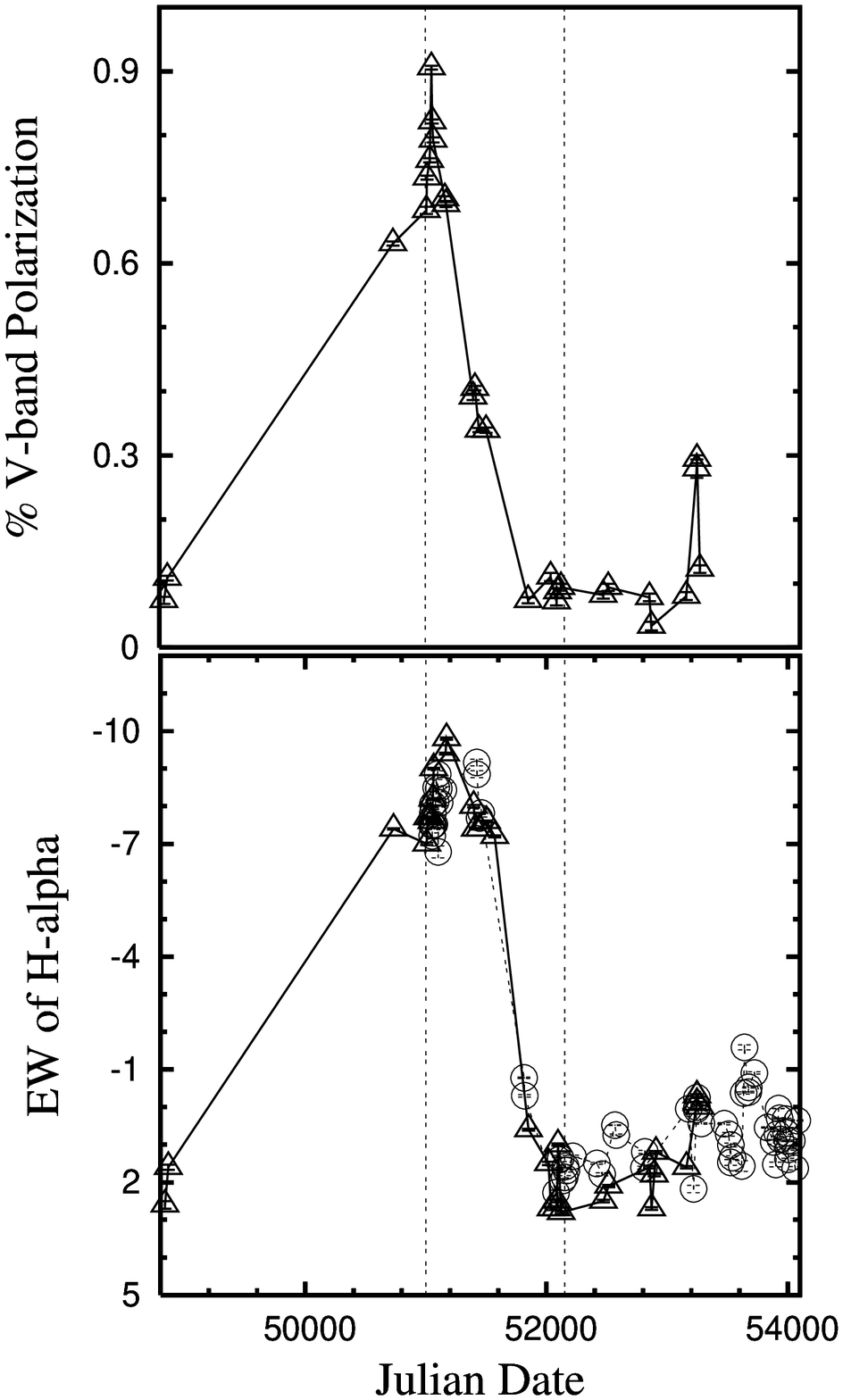}
\includegraphics[width=8cm]{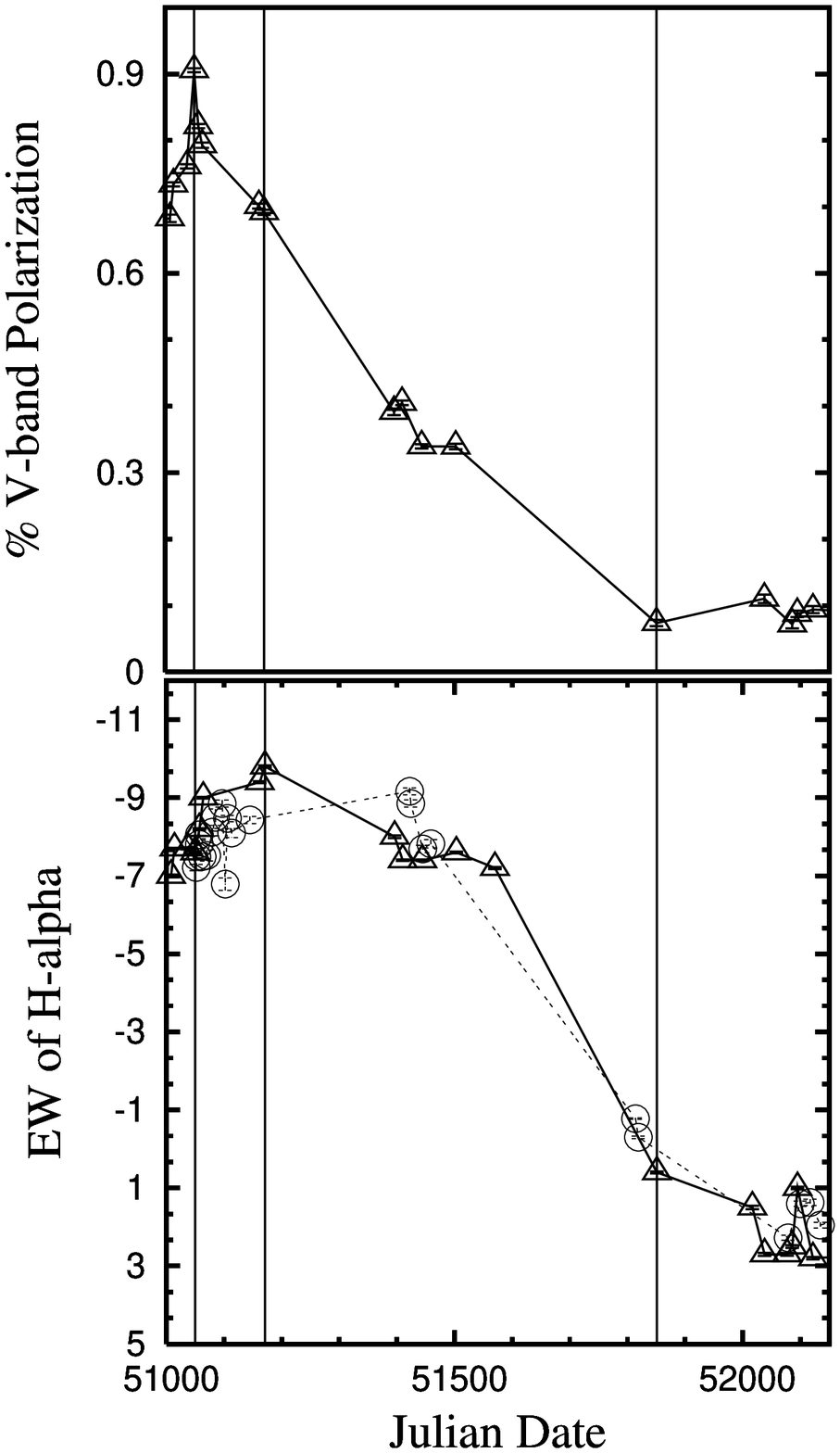}
\caption{The time evolution of 60 Cyg's observed (intrinsic and ISP) V-band polarization (top panels) and H$\alpha$ EW (bottom panels) is illustrated for our 
entire dataset (left figure set) and a zoomed region (right figure set).  Open circles correspond to Ritter observations while open triangles correspond to HPOL observations, and error bars (compiled in Table \ref{60cygdata}) are smaller than the size of the data points.  As discussed in Section \ref{data}, a factor of -1.9 \AA\ has been added to HPOL EW data compiled in Table \ref{60cygdata}.  The three vertical lines in the right figure set depict (from left to right) the epoch of the maximum V-band polarization (JD=2451049), the maximum H$\alpha$ EW (JD=2451171), and the minimum H$\alpha$ EW (JD=2452038).  The monotonically decreasing H$\alpha$ EW strength demonstrates the gradual loss of 60 Cyg's disk over a time-scale of $\sim$870 days.  The $\sim$120 day lag of 
the maximum H$\alpha$ EW from the maximum V-band polarization and the $\sim$190 day lag of the onset of the minimum H$\alpha$ EW from 
the onset of the minimum V-band polarization suggest that the disk-clearing progressed in an ``inside-out'' manner.  The full time-scale for disk loss, from the maximum V-band polarization to the minimum H$\alpha$ EW, is $\sim$1000 days ($\sim$870 days + $\sim$120 days). \label{60cygew} }
\end{center}
\end{figure}

\begin{figure}
\begin{center}
\includegraphics[width=8cm]{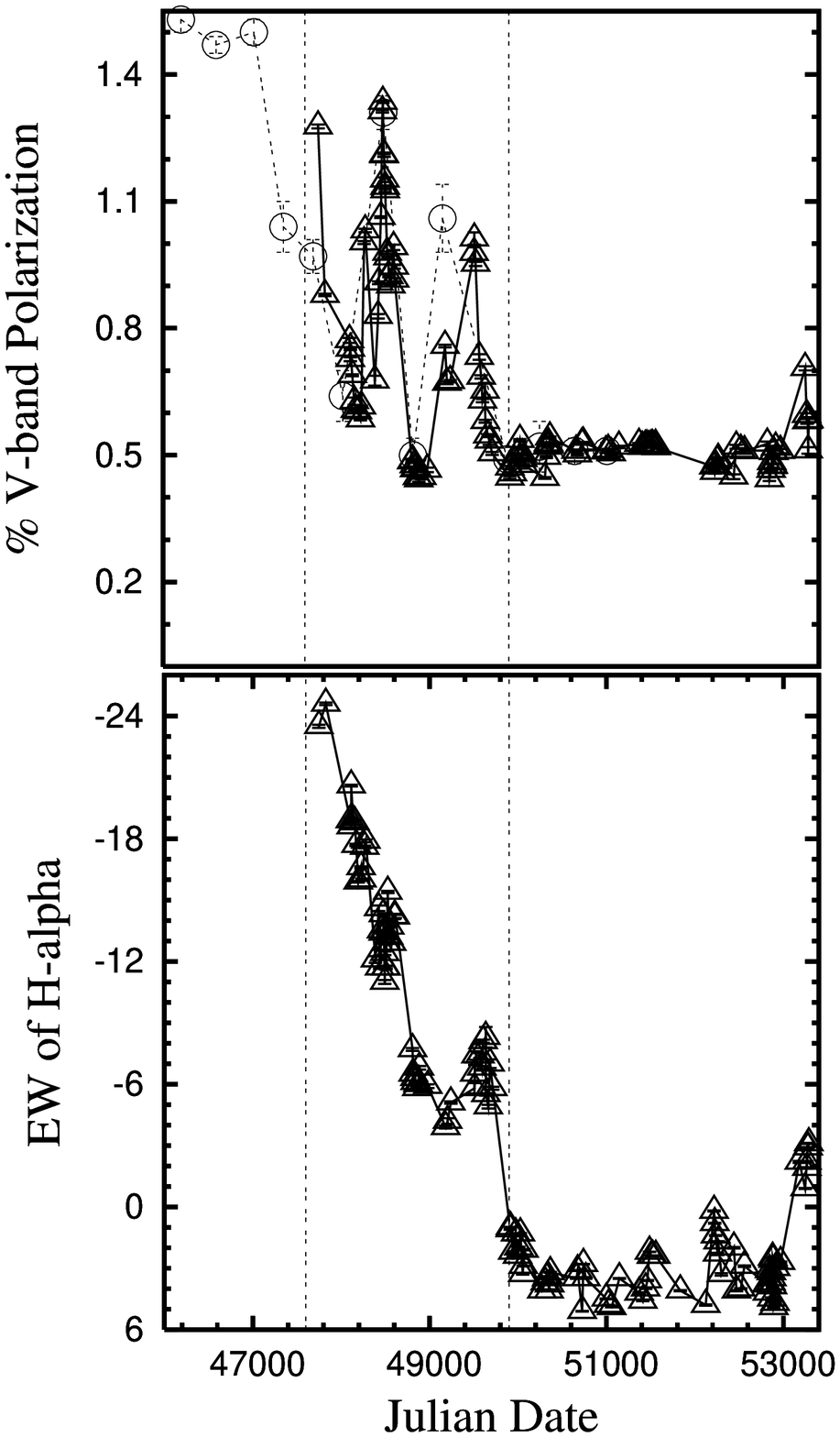}
\includegraphics[width=8cm]{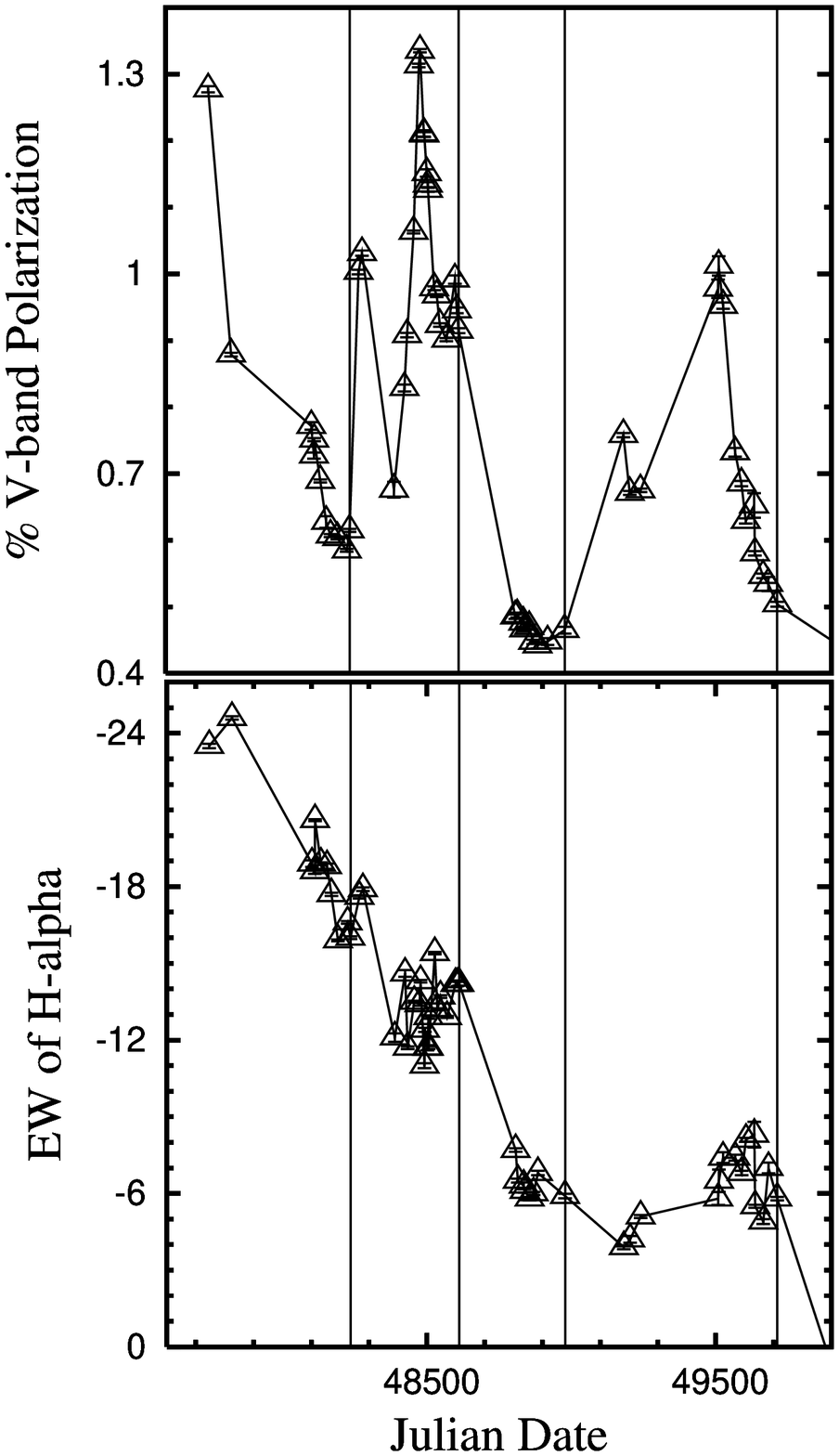}
\caption{The temporal evolution of $\pi$ Aquarii's observed (intrinsic and ISP) V-band polarization (top panels) and H$\alpha$ EW (bottom panels) are illustrated 
for our entire dataset (left figure set) and a zoomed region (right figure set).  Open triangles correspond to our HPOL data while open circles 
correspond to literature V-band polarization values compiled by \citet{mcd86,mcd90,mcd94,mcd99}.  HPOL EW and polarization errors (compiled in 
Table \ref{piaqrdata}) are smaller than the size of the data points.  The vertical lines in the right figure set denote 
(from left to right) the onset (JD=2448226) and termination (JD=2448612) of the $\sim$390 day long first polarimetric outburst and the onset (JD=2448979) and termination (JD=2449713) of the $\sim$730 day long second polarimetric outburst.  \label{piaqrew}}
\end{center}
\end{figure}

The time evolution of 60 Cyg's observed (intrinsic and ISP) V-band polarization is depicted in the top panels of Figure \ref{60cygew} and tabulated in Table \ref{60cygdata}.  While the observed magnitude of polarization mimics the similar long-term trend as the strength of the H$\alpha$ line 
emission, the maximum H$\alpha$ emission lags the peak polarization observed on 1998 August 24 by $\sim$120 days, as depicted by the first two vertical lines in the right panels of Figure \ref{60cygew}.  Similarly, the onset of the minimum H$\alpha$ EW (third vertical line, Figure \ref{60cygew}) lagged the onset of the 
polarization low-state by $\sim$190 days.  Such a time lag between the onset of polarimetric and spectroscopic features in Be stars has been 
previously reported in the literature (see e.g. \citealt{po78a,po78b}) and is indicative that the clearing of the disk proceeds in an 
 ``inside-out'' manner.  If we assume that the time period immediately following the polarization maximum denotes the beginning of the 
 disk-loss transformation, then the full time-scale for disk loss is simply the summation of the H$\alpha$ EW decline time-scale plus the 
 polarization time lag ($\sim$870 days + $\sim$120 days), $\sim$1000 days.  This disk loss time-scale is similar to the $\sim$700 day time-scale \citet{poe82} 
 noted for the loss of omicron And, using the same definition of the start and termination of the disk-loss phase.
 
 In Section \ref{60cyghalpha} we noted the presence of minor, temporary increases in the H$\alpha$ EW after 60 Cyg reached its diskless low state.  Inspection of Figure \ref{60cygew} illustrates that one of these EW events near 2004 Aug 30-31 was accompanied by a $\sim$0.3\% V-band polarimetric outburst, strengthening our interpretation that these minor outbursts represent tenuous attempts of the 
 system to reform a gaseous circumstellar disk.
 
 Though not plotted in Figure \ref{60cygew} for asthetic reasons, the PBO Lyot polarization observation from 1987 (Table \ref{lyot}), which encompasses 
 the V-band, is consistent with HPOL polarimetry obtained during epochs in which 60 Cyg was characterized by a strong disk.  These data are also consistent with the results of \citet{kou00}, who reported strong H$\alpha$ emission present before (early 1980s) and after (later 1980s) our PBO Lyot observation.

\subsubsection{$\pi$ Aqr} \label{piaqrpol}

The time evolution of $\pi$ Aqr's observed V-band polarization, depicted in the top panels of Figure \ref{piaqrew} and tabulated in Table 
\ref{piaqrdata}, exhibits a substantially more complex morphology during the star's disk-loss episode as compared to 60 Cyg.  Archival 
V-band polarization prior to the onset of our observations \citep{mcd86,mcd90}, plotted as open circles in Figure \ref{piaqrdata}, indicate 
that the disk was in a stable high state between 1985-1987.  Figure \ref{piaqrdata} clearly shows that the $\sim$2440 day long decline of the H$\alpha$ EW was interupted by two massive polarimetric outbursts.  The $\sim$390 day long first polarimetric event, demarked by the first and second vertical 
lines in the right panels of Figure \ref{piaqrdata}, is clearly related to the $\sim$220 day long first H$\alpha$ EW event 
noted in Section \ref{piaqrhalpha}.  The $\sim$730 day long second polarimetric event, demarked by the third and fourth vertical lines in the right panels 
of Figure \ref{piaqrdata}, is similarily related to the $\sim$200 day second interuption of the H$\alpha$ EW decline noted in Section \ref{piaqrhalpha}, during which the EW briefly began to re-strengthen.

Given the complex morphology of both the V-band polarization outbursts and the interruptions of the H$\alpha$ EW decline, it is more 
difficult to quantify the exact time lag between the onset of polarimetric features and the corresponding effects in H$\alpha$.  However, 
inspection of Figure \ref{piaqrdata} clearly shows that such a time lag does exist; for example, the total V-band polarization reaches its 
low state much earlier than the time at which the H$\alpha$ EW reaches its low state.  Similar to our interpretation of our 60 Cygni data, we 
deduce this time lag as evidence that $\pi$ Aqr's disk cleared in an ``inside-out'' fashion.

In Section \ref{piaqrhalpha} we noted the presence of two minor, temporary increases in the H$\alpha$ EW after the polarimetric data 
indicated that $\pi$ Aqr had reached its 
quiescent, diskless state.  We note that the final EW increase, which began in 2003 October and was still ongoing at the termination of our 
observational data in 2004 October, coincided with a corresponding temporary increase in the system's total V-band polarization, which 
peaked at 0.2\% above the background interstellar polarization level on 2004 August 29.  We observed no increase in the polarization  
corresponding to the 2001 November - 2002 June H$\alpha$ EW outburst, and speculate that we missed observing the onset of this event due to 
the sparse time sampling of our polarimetric observations during this epoch.

Though not plotted in Figure \ref{60cygew} for asthetic reasons, the PBO Lyot polarization observations from 1979-1980 (Table \ref{lyot}), which encompass the V-band, are consistent with HPOL polarimetry obtained during epochs in which $\pi$ Aqr was characterized by a strong disk and also 
consistent with the general level of V-band polarization in the late 1970s presented by \citet{bjo02}.

\subsection{Interstellar Polarization} \label{isp}

While polarimetry can be used as a unique diagnostic tool to investigate the properties of astrophysical objects, proper 
interpretation of these data is inherently difficult, as observations can contain contributions which are both intrinsic and
interstellar in origin.  The identification and complete removal of the interstellar polarization component (hereafter ISP), and a thorough 
understanding of the accuracy of this removal, is thus a critical first step in interpreting polarimetric data.  
As detailed by \citet{mcl78} and discussed in \citet{qui97}, there are three techniques commonly used to determine ISP along a line of sight: field star
studies, emission line polarization analysis, and wavelength dependence studies.  

\subsubsection{Field Star Technique} \label{fs}

Field star studies are the most basic of the ISP separation techniques, 
and due to their large uncertainties, they are most useful when used in
combination with other methods.  The technique involves averaging the observed polarization of stars 
at the similar distances to and small angular separations from the science target.  It inherently assumes the existence of
uniform interstellar medium conditions in the vicinity of the target and also assumes that the field stars 
themselves are devoid of any intrinsic polarization component.  
While the technique can be highly successful when used to analyze the ISP properties of open clusters, where one can 
assume all members are located at the same distance and there are sufficient numbers of stars to lessen the effects 
of small intrinsic polarization components in the field stars by brute force averaging of a large number of stars 
(see e.g. \citealt{wi07b}), the application of the technique to isolated stars often yields imprecise ISP estimates (see \citealt{mcl79}, 
\citealt{bjo98}, and \citealt{wis06}). 

We were able to identify seven non-emission line stars with polarimetric data in the catalog 
of \citet{mat70} spatially near the position of $\pi$ Aqr on the sky which appeared to be suitable field stars (see Table \ref{t2}).
Averaging these field stars yields the ISP parameters of $P_{ism} = 0.47\%$ at $\theta_{ism} = 130.1 ^{\circ}$ .  
As we will show in Section \ref{serk}, this initial ISP estimate does not fully remove the interstellar polarization component 
from our data; hence, we do not adopt these values as our final ISP determination along the line of sight to $\pi$ Aqr.

\begin{figure}
\begin{center}
\includegraphics[width=10cm]{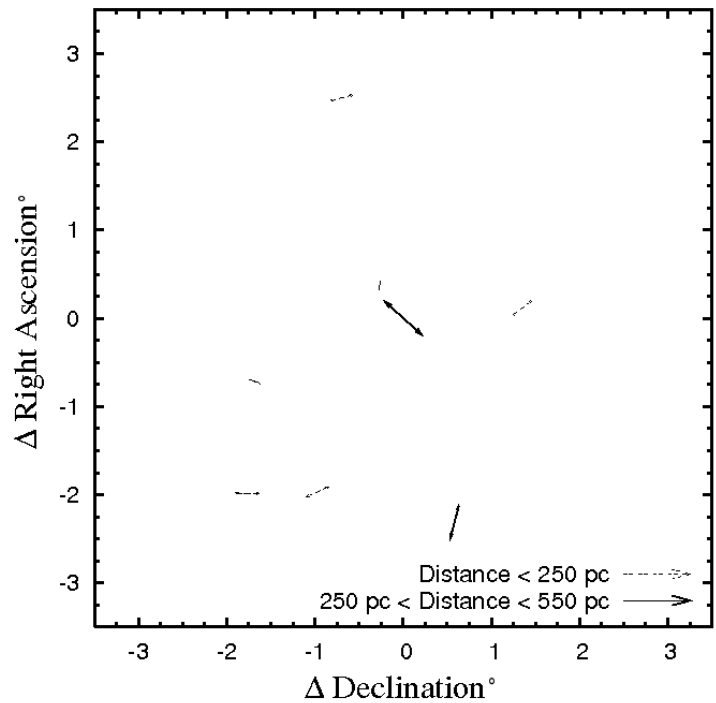}
\caption{The polarization values of stars nearby the location of 60 Cyg, based on the polarization catalog of \citet{hei00}.  One star having a similar distance (250 - 550 pc) as 
60 Cyg (418 pc \citealt{per97}), shown with a solid line, exhibits significantly different polarization than 60 Cyg, indicating that a useful ISP estimate along 
the line of sight to 60 Cyg cannot be obtained from these data.  For reference, the length of the vectors in the legend correspond to a polarization magnitude of 0.325\%.\label{60cygfield}}
\end{center}
\end{figure}

Using the polarization catalog of \citet{hei00}, we were only able to identify one non-emission line star with suitable polarimetric data having a similar distance as 
60 Cyg (418 pc, \citealt{per97}; see Table \ref{t2}).  As seen in Figure \ref{60cygfield}, most of the stars spatially 
coincident with 60 Cyg are in the foreground; moreover, the one star at a similar distance (250 - 550 pc) as 60 Cyg exhibits a significantly  
different polarization.  We conclude that the field star technique fails to yield a useful estimate of the ISP along the line of 
sight to 60 Cyg.

\subsubsection{H-$\alpha$ Line Depolarization} \label{line}

If one assumes that the physical location of the emitting region of emission lines in an astrophysical object lies exterior 
to the region producing any intrinsic polarization, as suggested
by \citet{har68}, one can measure the polarization across emission lines such as H$\alpha$ and use 
these values as measures of the ISP along the line of sight.  While this technique has been successfully employed 
for a wide variety of objects \citep{nor01,mey02,wis06}, \citet{mcl79} note that the semi-empirical relationship used 
to isolate ISP via this technique is questionable and subject to numerous complications.  \citet{qui97} remark that the technique's 
assumption of a constant intrinsic position angle across
H$\alpha$ is also doubtful and note that their own spectropolarimetric observations show H$\alpha$ is not completely
unpolarized.  \citet{poe79}, \citet{oud05}, and \citet{har09} also demonstrate that Be stars can exhibit intrinsic line polarization features.  The reliability of this technique is therefore uncertain at best.

To apply this technique to our data, we first created artificial filters which sampled regions within the H$\alpha$ line core (6520 -6620\AA) and 
continuum on both sides of the line (6350 -6450 \AA\ and 6700 -6800 \AA).  We then determined the polarization within the H$\alpha$ 
line and its nearby continuum, using these filters, for all data which showed clear spectroscopic evidence of having H$\alpha$ in emission.  
We plotted the continuum and line polarizations for each observation of $\pi$ Aqr and 60 Cyg on separate Stokes Q-U diagrams, and 
looked for the standard evidence of full line depolarization effects.  Careful inspection of these figures revealed that the lines connecting each 
continuum-line observation set pointed in a different direction and that the H$\alpha$ line polarization measurements were not clustered around one central location.  These trends indicated that intrinsic polarization within the line was likely contaminating our data, thereby 
precluding our use of H$\alpha$ line depolarization as an ISP constraint for both $\pi$ Aqr and 60 Cyg. 

\subsubsection{Serkowski Law Fitting} \label{serk}

The wavelength dependence of interstellar polarization can be expressed by the empirical 
Serkowski law \citep{ser75}, as modified by \citet{wil82}: 
 $P(\lambda) = P_{max}$ exp
$\left [-K \ln ^{2} (\lambda _{max}/\lambda) \right ]$.  
In the absence of additional constraints, one can attempt to determine the ISP along the line of sight 
to an object containing both interstellar and intrinsic polarization if one has \textit{a priori} knowledge of the 
position angle of the instrinsic polarization component (see e.g. \citealt{poe79}).  Our $\pi$ Aqr and 60 Cyg 
spectropolarimetric data are particularily unique in that they span an era in which each star enters a ``normal B star'' phase 
in which all observational evidence of the presence of a disk, including an intrinsic polarization component, had 
disappeared.  The remnant polarization in these low-state data can be assumed to be fully interstellar in origin, thereby 
providing us the opportunity to accurately fit their wavelength dependence using a modified Serkowski law.

\begin{figure}
\begin{center}
\includegraphics[width=12cm]{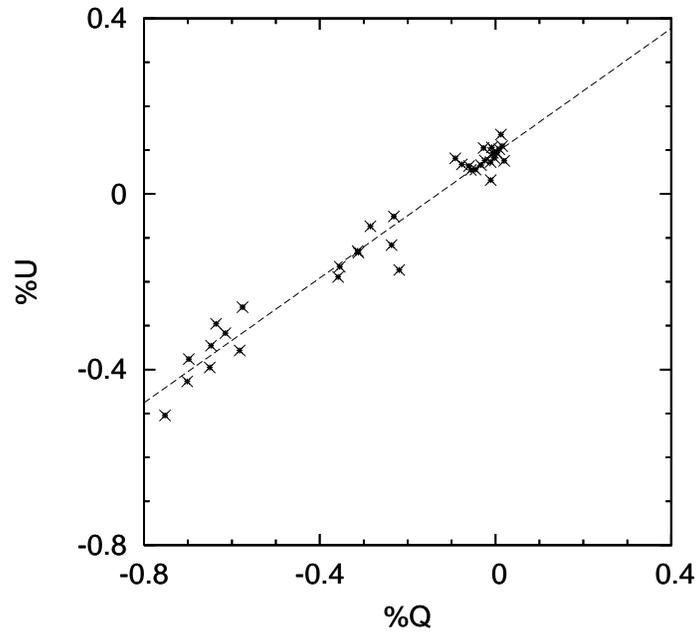}
\caption{The V-band polarization of our entire dataset for 60 Cyg is plotted on a Stokes Q-U diagram.  Note that the polarization errors 
(compiled in Table \ref{60cygdata}) are smaller than the size of the data points.  The data follow a strict linear trend indicating a constant scattering angle (i.e. polarization position angle) as expected in an electron scattering gaseous circumstellar disk system.  The slope of the best fit linear regression to the data, 0.71 $\pm$ 0.02, yields the intrinsic polarization position angle of 60 Cyg, $\theta_{\ast}$ = 107.7 $\pm$ 0.4$^{\circ}$.  This implies that 60 Cyg's disk is oriented at a position angle of $\theta_{disk}$ = 17.7$^{\circ}$ on the sky.   \label{60cygqu}}
\end{center}
\end{figure}

\begin{figure}
\begin{center}
\includegraphics[width=12cm]{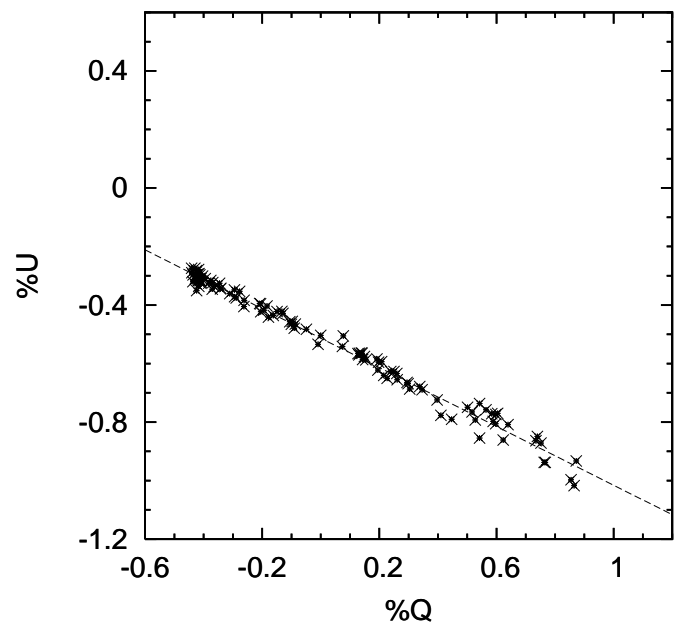}
\caption{The V-band polarization of our entire dataset for $\pi$ Aqr is plotted on a Stokes Q-U diagram.  Note that the polarization errors 
(compiled in Table \ref{piaqrdata}) are smaller than the size of the data points.  Similar to 60 Cyg 
(Figure \ref{60cygqu}), the data follow a linear trend indicative of a constant scattering angle as expected in a circumstellar disk 
system.  The slope of the best fit linear regression to the data, -0.50 $\pm$ 0.01, yields the intrinsic polarization position angle of 
$\pi$ Aqr, $\theta_{\ast}$ = 166.7 $\pm$ 0.1$^{\circ}$.  This implies that $\pi$ Aqr's disk is oriented at a position angle of 
$\theta_{disk}$ = 76.7$^{\circ}$ on the sky.  \label{piaqrqu}}
\end{center}
\end{figure}

We begin by plotting our entire V-band datasets for 60 Cyg and $\pi$ Aqr on a Stokes Q-U diagram as seen in Figures 
\ref{60cygqu} and \ref{piaqrqu} repectively, where 
\begin{equation}
Q = P cos (2\theta)
\end{equation}
\begin{equation} U = P sin (2\theta)
\end{equation}
\begin{equation}
 P = (Q^{2} + U^{2})^{0.5}
\end{equation}
\begin{equation}
\theta = 0.5 tan^{-1}(U/Q)
\end{equation}  The 
time evolution of the polarimetric data in each of these 
figures follows a strict linear trend, indicating a constant scattering angle (i.e. position angle) regardless of the magnitude 
of polarization, as is expected for electron scattering in a gaseous circumstellar disk.  The slope of the best fit linear regression to our 
60 Cyg data (Figure \ref{60cygqu}) was 0.71 $\pm$0.02.  This indicates that the intrinsic polarization position angle for 60 
Cyg is $\theta_{\ast}$ = 107.7 $\pm$0.4$^{\circ}$, and under the typical assumption that the maximum polarization occurs for 90$^{\circ}$ 
scattering it also implies that 60 Cyg's disk is oriented at a position angle of $\theta_{disk}$ = 17.7$^{\circ}$ on the sky (measured in the 
convention of North to East).  The 
slope of the best fit linear regression to our $\pi$ Aqr data (Figure \ref{piaqrqu}) was -0.50 $\pm$0.01.  This translates into an intrinsic 
polarization position angle, $\theta_{\ast}$ of 166.7 $\pm$0.1$^{\circ}$ for $\pi$ Aqr, and implies that $\pi$ Aqr's disk is oriented at a position 
angle of $\theta_{disk}$ = 76.7$^{\circ}$ on the sky.

Using an iterative process, we next carefully examined the polarimetric data which corresponded to time-periods in which H$\alpha$ was nearly or completely in absorption to help identify all observations which were devoid of an intrinsic polarization component.  Data comprised 
solely of an interstellar polarization component, which we labeled as ``L'' for \textit{polarization} low-state in Tables \ref{60cygdata} and \ref{piaqrdata}, clearly correlated with one another in a clump on the Stokes Q-U diagram (see e.g. Q $\sim$0\% and U$\sim$0.1\% in Figure \ref{60cygqu}).  Recall from Section 3.2 that H-$\alpha$ emission probes a much more extended region of a Be star's circumstellar environment 
than optical polarization measurements; therefore, our definition of a \textit{polarization} low-state only establishes that the density of material in the inner disk region dramatically dropped (or was completly cleared) and therefore does not demand that the outer circumstellar environment be similarly completely cleared of disk material.  Since the intrinsic polarization arising from classical Be stars follows a tight linear trend, we can effectively isolate the presence of any residual intrinsic polarization in our polarization low-state data by rotating all of our observations by the negative value of the intrinsic polarization position angle.  

\begin{figure}
\begin{center}
\includegraphics[width=15cm]{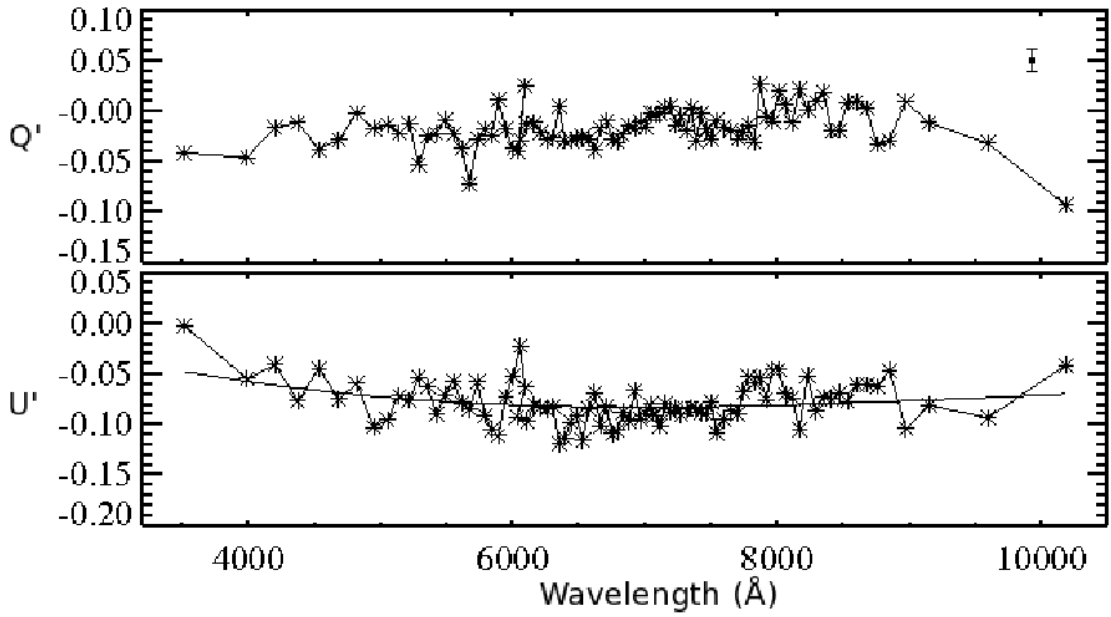}
\caption{The wavelength dependence of 60 Cyg's polarization low-state data, binned to a constant (Poisson photon statistic) polarization error of 0.01\% (depicted as a representative error bar), in the rotated Stokes U$^{'}$ component (see Section \ref{serk}) was fit with 
a modified Serkowski law \citep{ser75,wil82}, yielding the estimate of the perpendicular ISP component along the line of sight to the star.  The data in the rotated Stokes Q$^{'}$ component, representing the parallel component of the ISP, was close 
to zero and not well fit by a Serkowski law.  Section \ref{serk} provides a discussion of how we estimated the parallel ISP component for 60 
Cyg.  The parallel, perpendicular, and total ISP values adopted in this paper are summarized in Table \ref{isptable}.  \label{60serk}}
\end{center}
\end{figure}

\begin{figure}
\begin{center}
\includegraphics[width=12cm]{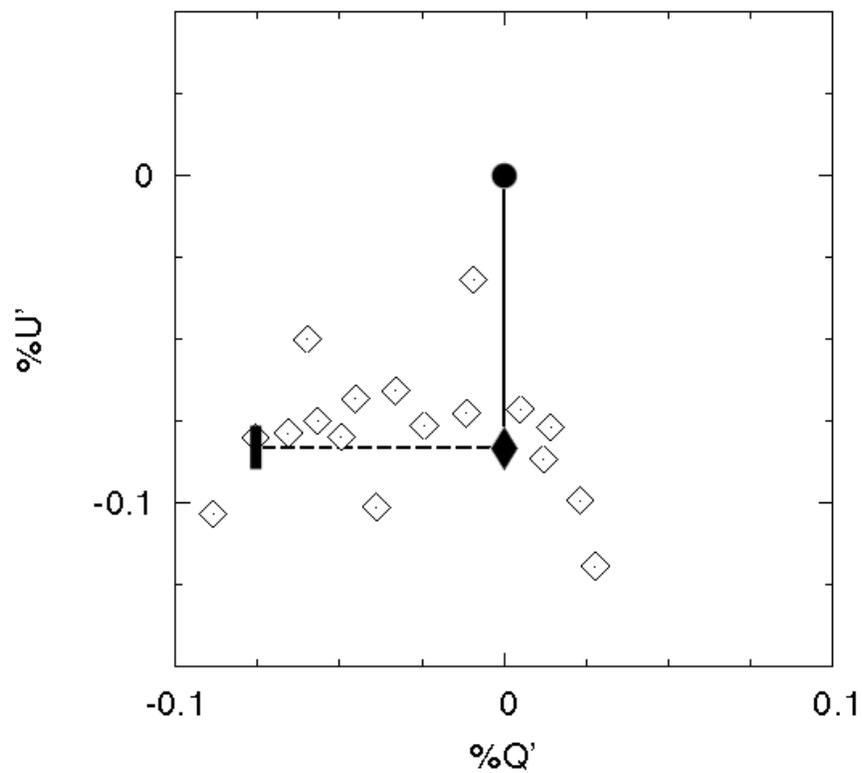}
\caption{The low state (diskless) 60 Cyg V-band polarimetric data are plotted on a Stokes Q$^{'}$-U$^{'}$ diagram.  Errors are smaller than the 
size of the data points.  As discussed in 
Section \ref{serk} and illustrated in Figure \ref{60serk}, we fit a modified Serkowski law to the U$^{'}$ component of these data to 
extract the perpendicular component of the interstellar polarization (denoted here by the solid line).  Since the wavelength dependence of 
the U$^{'}$ data were not well represented by a Serkowski law (see Figure \ref{60serk}, we determined the minimum parallel component of 
the interstellar polarization to be simply the vector along the Q$^{'}$ direction which joined the end of the perpendicular ISP component with 
the data having the minumum Q$^{'}$ value, (denoted by the dashed line).  This minimum parallel ISP component is tabulated in Table \ref{isptable}. \label{60cygrotqu}}
\end{center}
\end{figure}

\begin{figure}
\begin{center}
\includegraphics[width=12cm]{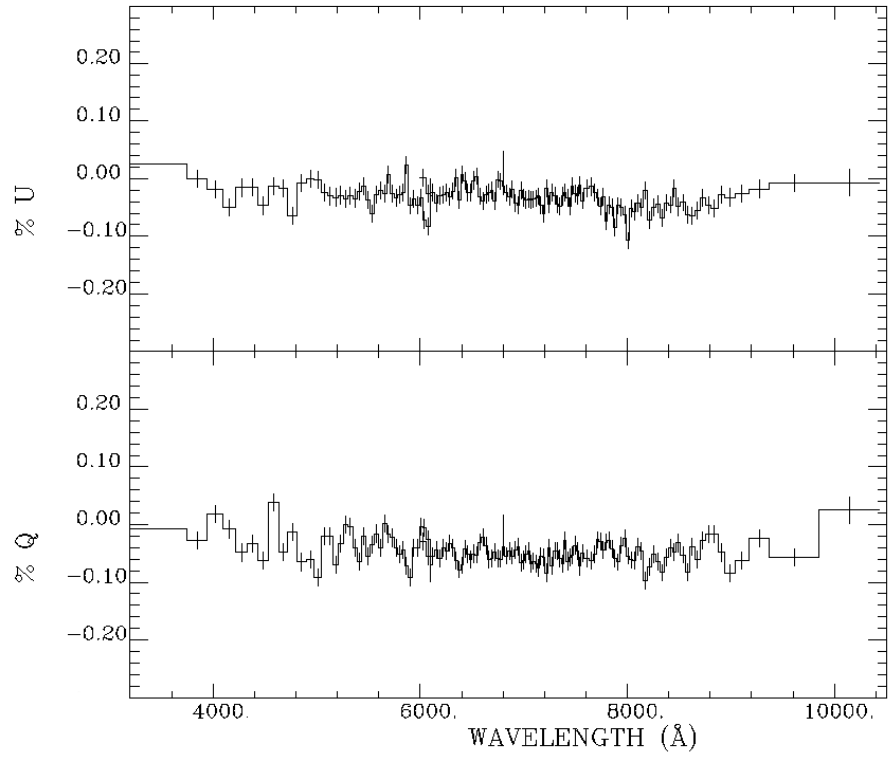}
\caption{The residuals of our low-state 60 Cyg data, binned to a constant (Poisson photon statistic) polarization error of 0.015\%, are shown after 
subtracting the total ISP values listed in Table \ref{isptable}.  The data are consistent with Q = 0\% and U = 0\% and exhibit no 
significant wavelength dependence, providing confidence in the total interstellar polarization values we derived 
for 60 Cyg. \label{60cygresid}}
\end{center}
\end{figure}

For 60 Cyg, we rotated our data by -107.7$^{\circ}$, thereby placing any small amount of residual intrinsic polarization in the Stokes Q$^{'}$ parameter (where the prime denotes the rotated frame) and fully isolating the perpendicular component of 
the interstellar polarization in the Stokes U$^{'}$ parameter.  After co-adding all low state polarization data in this rotated frame, we then fit a modified 
Serkowski law to the U$^{'}$ data using the IDL subroutine \textit{mpcurvefit}, as demonstrated in Figure \ref{60serk}, which plots the U$^{'}$ component of the 60 Cyg data binned to a constant (Poisson photon statistic) polarization error of 0.01\%.  60 Cyg's perpendicular ISP component (U$^{'}$ data) is clearly well fit by a modified Serkowski law having parameters listed in Table \ref{isptable}.  There is very little signal in 60 Cyg's Q$^{'}$ component (Figure \ref{60cygqu}); hence, we 
could not fit the data with a modified Serkowski law to extract the parallel ISP component.  Instead, we overplotted our perpendicular ISP component on our low-state polarization data in the Stokes Q$^{'}$-U$^{'}$ parameter space and estimated the lower-limit of the parallel ISP component 
needed to ensure that our lowest state datapoint was purely interstellar in nature (see dashed line in Figure \ref{60cygrotqu}).  
Our lower limit estimate of the parallel ISP is tabulated 
in Table \ref{isptable}, along with the total ISP estimate derived from vectorally adding the perpendicular and parallel ISP components.  We 
find the total ISP for 60 Cyg to be characterized by P$_{max}$ = 0.112\% $\pm$ 0.001\%, $\lambda_{max}$ = 6977 $\pm$ 241\AA, PA = 41.5$^{\circ}$, and 
K = 1.17.  The wavelength dependence of our polarization low state 60 Cyg data after subtraction of this total ISP, seen in Figure \ref{60cygresid} binned 
to a constant polarization error of 0.015\%, is centered about Q = 0\% and U = 0\% and exhibits no significant structure, thereby confirming 
the validity of our ISP determination.

\begin{figure}
\begin{center}
\includegraphics[width=15cm]{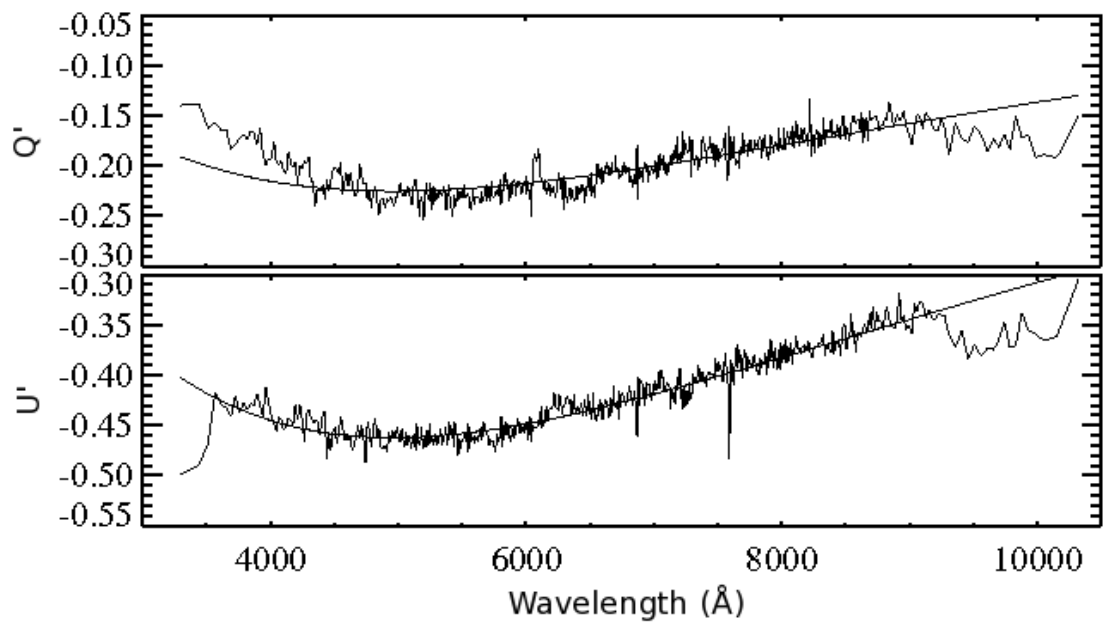}
\caption{The wavelength dependence of $\pi$ Aqr's polarization low-state data, binned to a constant (Poisson photon statistic) polarization error of 0.01\%, in the rotated Stokes U$^{'}$ component (see Section \ref{serk}) was fit with 
a modified Serkowski law \citep{ser75,wil82}, yielding the estimate of the perpendicular ISP component along the line of sight to the star (shown).  
The data in the rotated Stokes Q$^{'}$ component, representing the parallel component of the ISP, were 
also fit with a modified Serkowski law (shown), which included an additional constant term to account for the presence of a potential remnant 
intrinsic polarization component.  The parallel, perpendicular, and total ISP values adopted in this paper are summarized in Table \ref{isptable}.    \label{piaqrserk}}
\end{center}
\end{figure}

We employed a similar technique to determine the ISP for $\pi$ Aqr.  We first rotated our data by -166.7$^{\circ}$ to place any small level of 
residual intrinsic polarization in our low-state data in the Stokes Q$^{'}$ parameter and isolating the perpendicular ISP component in the 
Stokes U$^{'}$ parameter.  We were able to obtain a good fit of a modified Serkowski law to the co-added U$^{'}$ data again using \textit{mpcurvefit}, as demonstrated in Figure \ref{piaqrserk}, which plots $\pi$ Aqr's U$^{'}$ component binned to a constant (Poisson photon statistic) polarization 
error of 0.01\%.  We do note that a small fraction of the binned U$^{'}$ data at the extrema of our wavelength coverage, e.g. below $\sim$4000 \AA\ and 
above $\sim$9500\AA, deviate from the best fit by 
$\le$0.1\%.  These deviations are less than 3-$\sigma$ from the known instrumental systematic uncertainties (Section \ref{data}); hence, we do not 
believe they represent real features.   To confirm this, we also fit the data with an error weighted modified Serkowski-law, after adopting the known 
systematic instrumental uncertainties for the U- and I-band data, which were larger than the Poisson photon statistic errors.  Similar to 
the result achieved for completely excluding the U- and I-band data from the fit, we found this exercise did not appreciably affect the resultant best fit ISP parameters.  Since the number of binned data points which deviate from the fit is extremely small, relative to the total number of binned data points used the fit, these two modifications to the fitting routine had no effect on the resultant best fit.  The Serkowski law parameters of $\pi$ Aqr's perpendicular ISP component are listed in Table \ref{isptable}.  

\begin{figure}
\begin{center}
\includegraphics[width=10cm]{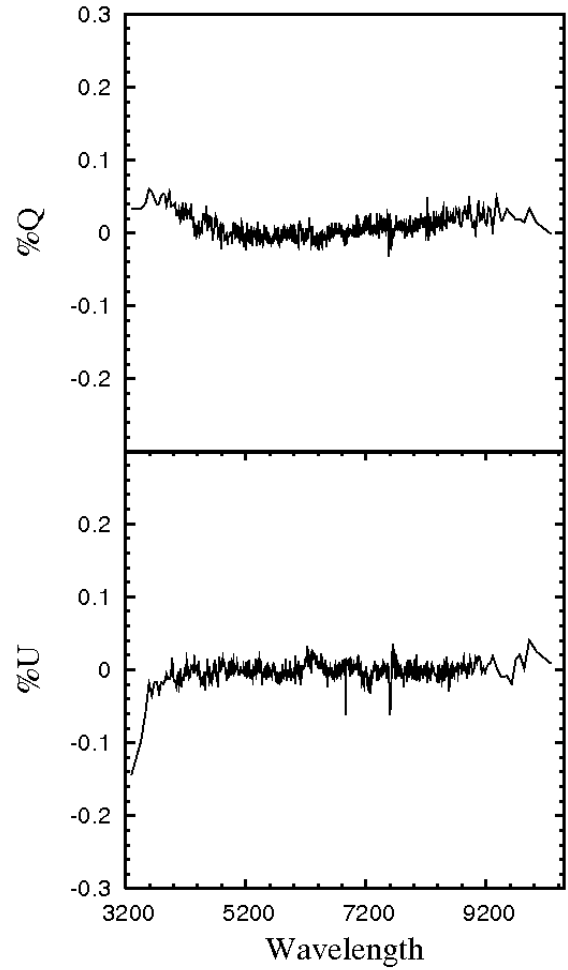}
\caption{The residuals of our polarization low-state $\pi$ Aqr data, binned to a constant (Poisson photon statistic) polarization error of 0.01\%, are shown after 
subtracting the total ISP values listed in Table \ref{isptable}.  The data exhibit a low level of residual polarization ($<$ 0.05\%) in both 
the Stokes Q and U parameters and exhibits little structure outside of the regions in which the HPOL CCD sensitivity is low ($<$ $\sim$4000 \AA\ and $>$ $\sim$9500 \AA, providing confidence in the total interstellar polarization values we derived for $\pi$ Aqr.
  \label{piaqrresid}}
\end{center}
\end{figure}

For the parallel ISP component, we included an additional constant term in our modified Serkowski law fitting, ``C'', to accomodate the possibility that a 
small level of intrinsic polarization was present in our low state data and fixed the value of $\lambda_{max}$ to that found in the 
perpendicular ISP component (Table \ref{isptable}), as this parameter should be the same for both the perpendicular and parallel ISP components.  The resultant best fit to the Q$^{'}$ component agrees reasonably well with the data, 
as seen in Figure \ref{piaqrserk}, with small deviations present below $\sim$4000 \AA\ and above $\sim$9500\AA.  These $\leq$0.05\% deviations 
from the best fit are less than 3-$\sigma$ from the known instrumental systematic uncertainties (Section \ref{data}); hence, we do not believe they are 
real and, as for the perpendicular ISP fit, they do not significantly influence the best fit parallel ISP parameters.  We find the total ISP for $\pi$ Aqr to be characterized by P$_{max}$ = 0.514\% $\pm$ 0.001\%, 
$\lambda_{max}$ = 4959 $\pm$15 \AA, PA = 108.7$^{\circ}$, and K = 0.83, as computed by vectorally adding the parallel and perpendicular 
ISP components listed in Table \ref{isptable}.  The wavelength dependence of our polarization low state $\pi$ Aqr data after subtraction of this total ISP, 
seen in Figure \ref{piaqrresid} binned to a constant polarization error of 0.005\%, exhibits only a small amount of residual polarization ($<$ 0.1\%) in either the Stokes Q or U components and exhibits little structure outside of the aforementioned regions in which the HPOL CCD sensitivity is low.

\subsubsection{Comparison to Previous ISP Estimates}

While we are not aware of any successful attempts to characterize the ISP along the line of sight to 60 Cyg 
(see e.g. \citealt{poe79}), two groups of authors have attempted to derive the ISP along the line of sight to 
$\pi$ Aqr.  \citet{mcl79} estimated the ISP to be $P_{ism} \sim
0.45 \%$ and $\theta_{ism} \sim 130^{\circ}$ from analysis of field stars, $P_{ism} \ge 0.6
\%$ based upon $\pi$ Aqr's E(B-V) excess, and $P_{ism} \le 0.4 \%$ and $\theta_{ism} \le
130^{\circ}$ from studying the wavelength dependence of broad-band observations.  The authors also note that 
the latter parameters were supported by H$\alpha$ and H$\beta$ line polarimetry.  From these various methods, 
\citet{mcl79} adopted the final ISP values of $P_{max} \sim 0.36 \%$, $\lambda_{max} = 5400$\AA , $\theta_{ism}
= 120^{\circ}$, and placed no formal errors on these estimates.  \citet{poe79} analyzed a heterogeneous mix of optical 
intermediate-band filter polarimetry, H$\alpha$ polarimetry, and IR polarimetry to derive three different ISP estimates with 
large reduced $\chi^{2}$ values ranging from 5.6-10.7.  While the $\lambda_{max}$ value derived by \citet{poe79} agrees 
with the value we have derived (Table \ref{isptable}) their PA and P$_{max}$ values differ by $\sim$7$^{\circ}$ and up to 0.1\% 
respectively.

\section{Discussion} \label{discussion}

\subsection{Disk-Loss and Disk-Renewal Time-scales}

Identifying the mechanisms responsible for the disk-loss and disk-renewal phases of classical Be stars (i.e. the transition to and from a ``Be'' phase and a normal B-star phase) can be aided by an examination of the time-scale of some or all of the four main phases involved: the disk growth phase, the steady-state disk phase, the disk-loss phase, and the steady-state quiescent (diskless) phase.  

Our data poorly constrain the time-scale for the complete regeneration of an active disk state from a quiescent diskless state.  Owing to the poor temporal sampling of our 60 Cyg data between 1992 and 1997 (Figure \ref{60cygew}), we can only place an upper limit on the regeneration of the active disk state observed in our data of $<$5 years.  We do see evidence of temporary, correlated enhancements in both the V-band polarization and H$\alpha$ EW during 60 Cyg's quiescent, diskless state, including one moderate event which occurred between June and September 2004 in the 
V-band polarization data.  Similarily, our long-term observations of $\pi$ Aqr do not cover a full disk regeneration cycle which would enable us to constrain this time-scale.  However, we do observe one major, correlated enhancement in both the V-band polarization and H$\alpha$ EW during the quiescent phase (Figure \ref{piaqrew}), again during the June - September 2004 time-frame in the V-band polarization data.  The sparse temporal sampling of 60 Cyg's and $\pi$ Aqr's outburst episodes during their quiescent phases only enables us to place upper limits on the time-scales of their duration of $<$ a few months.  

H$\alpha$ monitoring of the B3 star $\mu$ Cen, revealed similar types of outburst events occuring during the star's quiescent phase, which 
\citet{han93} interpreted as evidence of individual blobs of material being injected into the star's inner circumstellar environment on time-scales of 2-5 days, and then viscously accreting back onto the star and/or decreting on time-scales of 20-80 days.  Given the similarity in time-scales between $\mu$ Cen's quiescent-state H$\alpha$ outburst events and the duration of the polarization component of quiescent state outbursts observed 
in 60 Cyg and $\pi$ Aqr, we propose that our data might be diagnosing one (or several) injection events which subsequently viscously dissipate, perhaps arising from the same type of non-radial pulsation events thought to drive the injection events on $\mu$ Cen \citep{riv98}.  Such an interpretation is 
consistent with previous suggestions of non-radial pulsations in both $\pi$ Aqr \citep{pet05} and 60 Cyg \citep{kou00}.

As discussed in Sections \ref{60cyghalpha} and \ref{piaqrhalpha}, our data provide excellent temporal coverage of the disk-loss events of both 60 Cyg 
and $\pi$ Aqr.  60 Cyg experienced a generally monotonic reduction in H$\alpha$ emission strength from its maximum to minimum state over a 
time-scale of $\sim$870 days.  We interpret the $\sim$120 day lag of the maximum in the H$\alpha$ EW from the maximum V-band polarization as evidence that the 
disk cleared in an ``inside-out'' manner.  The total dissipation time-scale of 60 Cyg's disk, e.g. from the maximum polarization level to the 
minimum H$\alpha$ EW level, of $\sim$1000 days is similar to that observed for the dissipation of omicron And's disk ($\sim$700 days; \citealt{cla03}).  \citet{cla03} noted that omicron And's disk dissipation was consistent with the 
expected viscous time-scale of the disk, assuming the viscosity parameter $\alpha$ \citep{sha73} is $\sim$0.1.  Assuming that 60 Cyg's disk also 
dissipated on a viscous time-scale, we can derive the viscosity parameter $\alpha$ using equation 19 of \citet{bjo05}: 
\begin{equation}
t_{diffusion} = (0.2 yr / \alpha) * (r/R_{\ast})^{0.5}
\end{equation}.  Assuming that the bulk of the scattering events producing the observed polarization occurs at a radial distance of $\sim$2.5 R$_{\ast}$ \citep{car09} and a $\sim$800 day time-scale for the polarization to transform from maximum to minimum state (see Figure \ref{60cygew}), we derive a viscosity parameter of $\sim$0.14 for 60 Cyg, which is similar to that found for omicron And.  
By contrast, the loss of $\pi$ Aqr's disk as measured by its H$\alpha$ EW proceeded a factor of at least 2.4 times slower (e.g. $\sim$2440 days) than 60 Cyg's disk loss, and Figure \ref{piaqrew} clearly 
illustrates the presence of two extended ($\sim$390 day and $\sim$730 day) outburst events which temporarily stalled the long-term disk loss event.  Like 60 Cyg, $\pi$ Aqr also exhibited a lag between onset of the low-state of the V-band polarization and the low-state of the H$\alpha$ EW, indicating the disk 
cleared in an ``inside-out'' manner.  The exquisite temporal sampling of our $\pi$ Aqr data suggests that temporary increases in 
the mass injection rate into the inner disk may produce substantial deviations from a monotonic, viscous dissipation of Be disks.  We therefore suggest 
that claims of inconsistencies in the formation versus dissipation time-scales of disks with a standard viscous disk model based on sparsely 
sampled data, as made by \citet{vin06}, might be caused by under-sampling the type of short-lived outbursts observed in our $\pi$ Aqr data.

\subsection{Temporal Changes in the Intrinsic Polarization Position Angle} \label{pachanges}

\begin{figure}
\begin{center}
\includegraphics[width=10cm]{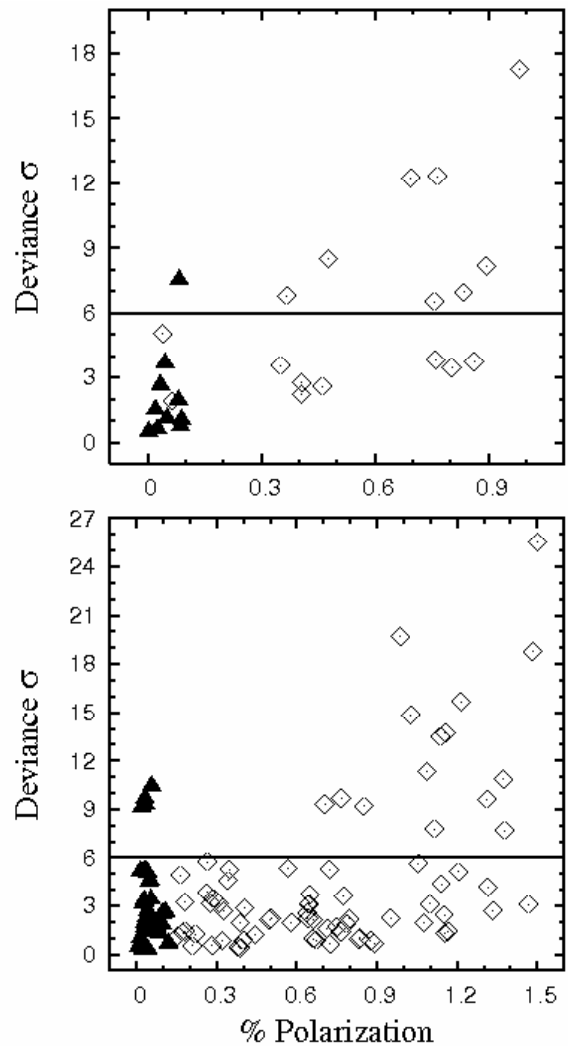}
\caption{The absolute value of the error weighted deviation (deviance) of every V-band polarization observation of 60 Cyg (top panel) and $\pi$ Aqr (bottom panel) from 
the best fit line in each star's Q-U diagram (Figures \ref{piaqrqu} and \ref{60cygqu}) is plotted as a function 
of the magnitude of intrinsic polarization present in each observation.  Filled triangles correspond to observations during the low (diskless) state 
of each star, while open diamonds correspond to observations during the high (disk present) state.  For both 60 Cyg and $\pi$ Aqr, the likelihood of 
observing a small-scale deviation in the polarization PA (from the bulk intrinsic PA of the disk) clearly increases with the magnitude of intrinsic polarization 
present.  A simple false discovery rate analysis (see Section \ref{pachanges}) confirms that this trend is statistically significant. \label{var}}
\end{center}
\end{figure}

While our total V-band polarization data for both 60 Cyg and $\pi$ Aqr both clearly followed a linear trend when plotted on a Stokes Q-U 
diagram (Figures \ref{60cygqu} and \ref{piaqrqu}), close inspection of these figures reveals tentative evidence of small-scale deviations in some of the data from each star's intrinsic polarization position angle, which seems to increase with the strength of the intrinsic polarization 
component.  To quantify this effect, we calculated the absolute value of the error weighted deviance of each data point from the best fit line denoting the star's intrinsic PA.  We plot these values as a function of the magnitude of the intrinsic polarization component for both 60 Cyg (top panel) and $\pi$ Aqr (bottom panel) in Figure \ref{var}, where filled triangles represent polarization low-state data (Tables \ref{60cygdata} and \ref{piaqrdata}) and 
open diamonds represent high-state data.  

Several trends are immediately apparent.  For both 60 Cyg and $\pi$ Aqr, the bulk of the data exhibit a $<$ 6$\sigma$ residual from the bulk value of the intrinsic PA value; moreover, very few data points at low intrinsic 
polarization levels deviate at the $>$ 6$\sigma$ level.  Just 2 of 15 (13\%) of 60 Cyg data below an intrinsic polarization level of 0.3\% 
exhibit a $>$ 6$\sigma$ deviation while a mere 4 of 59 (7\%) of $\pi$ Aqr data below an intrinsic polarization level of 0.6\% exhibit a 
$>$ 6$\sigma$ deviation.  By contrast, a substantial, statistically significant number of data at larger intrinsic polarization levels exhibit large deviations from the bulk intrinsic PA.  8 of 15 (53\%) of 60 Cyg data above an intrinsic polarization level of 0.3\% exhibit a 
$>$ 6$\sigma$ deviance while 15 of 32 (47\%) of $\pi$ Aqr data above an intrinsic polarization level of 0.6\% exhibit a $>$ 6$\sigma$ 
deviance.  If these trends are in fact real, they would indicate that disk material has a stronger likelihood to be located slightly 
above or below the disk midplane when the inner disk density is larger (i.e. when the intrinsic polarization component is larger).  For example, 
the position angle of the $\pi$ Aqr data point exhibiting a deviance of $\sim$26$\sigma$ deviated by 2.43$^{\circ}$ from the bulk intrinsic PA of the disk of $\theta_{\ast}$ of 166.7$^{\circ}$.

Before continuing our interpretation of this phenomenon, we consider the appropriateness of our choice of a deviance cutoff, above which we 
believe the data exhibit a bona-fide deviation from the intrinsic polarization PA (e.g. 6$\sigma$ in Figure \ref{var}).  Merely 
setting a threshold deviance value above which everything is considered real could lead to several false discoveries, owing to random 
variations in the data.  The false-discovery rate (FDR) method, developed by \citet{mil01} and employed by \citet{hop02} and \citet{kow09}, 
allows us a means to quantify the percentage of false positives above our choice of a threshold deviance value.  We model the null 
distribution (random noise) using the deviance values below an intrinsic polarization level of 0.6\% for $\pi$ Aqr and 0.3\% for 60 
Cyg.  Employing the IDL routine in Appendix B of \citet{mil01}, we determined that setting a threshold to the deviance value of 
7.687 for $\pi$ Aqr predicts a false-discovery rate of 32\%.  In other words, of the 19 discoveries above a deviance of 7.687 (spanning 
the entire intrinsic polarization range in Figure \ref{var}), no more than 32\% are caused by spurious deviations.  If we assume that the four 
discoveries below an intrinsic polarization of 0.6\% are in fact false discoveries, this implies that only 2 of the discoveries above 0.6\% are false, e.g. 13 of the 15 data points above the 6$\sigma$ level in the bottom panel of Figure \ref{var} are real.  Even in the worst case scenario 
that the entire 32\% of the sample of false discoveries resides in the discoveries we claimed above the 0.6\% intrinsic polarization level, a full 
68\% of this sample would still be real, statistically sound detections of deviations from the bulk polarization position angle.  Similarily, for 60 Cyg, a 
false-discovery rate of $\le$ 33\% sets the threshold deviance value at $\ge$ 6.52; hence, of the 9 observed values above this deviance level, at most 
3 could be false discoveries.  Based on this FDR analysis, we conclude most of our claimed deviations from the bulk PA of both 60 Cyg and $\pi$ Aqr are in fact real.

\begin{figure}
\begin{center}
\includegraphics[width=8cm]{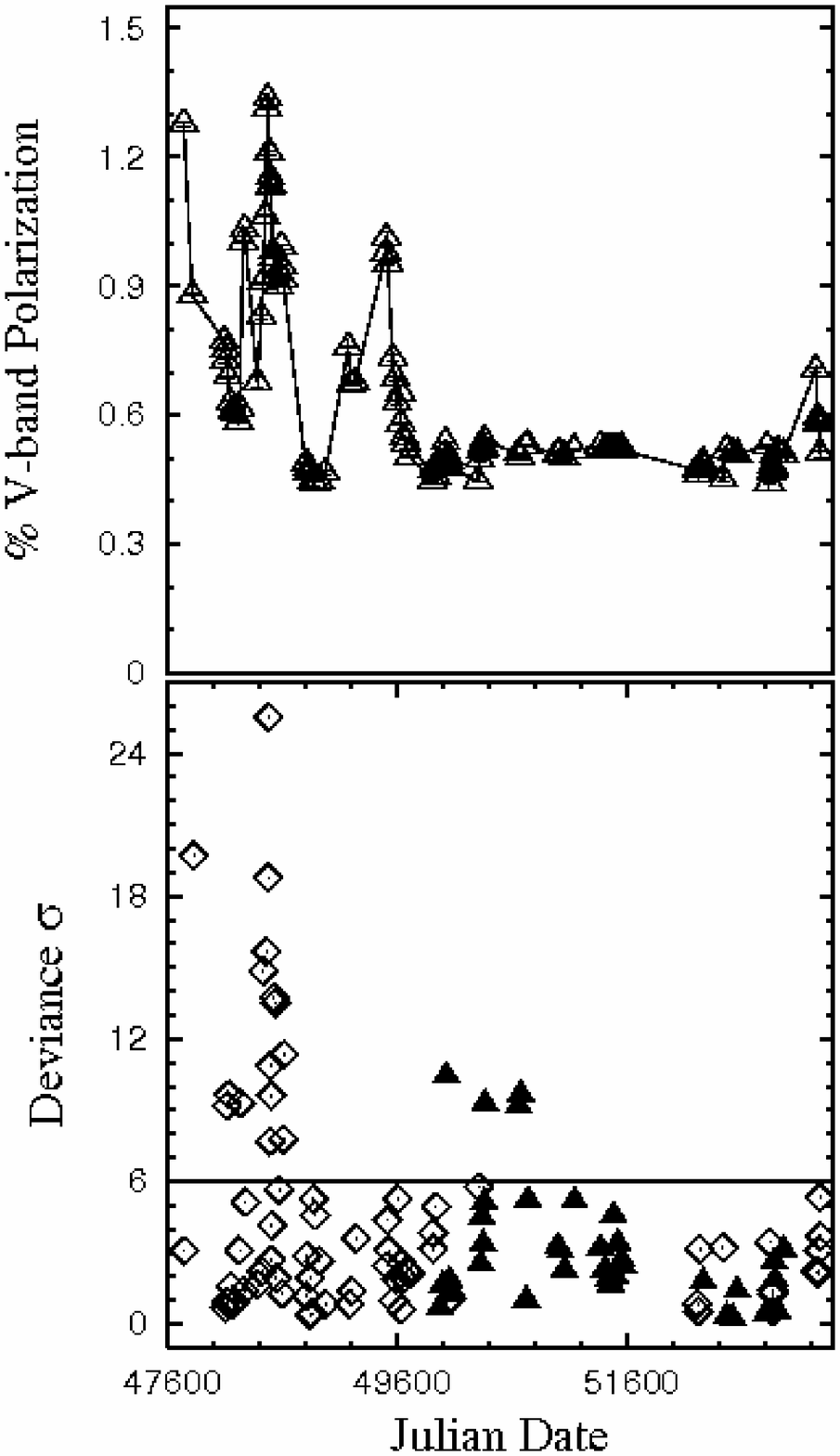}
\includegraphics[width=8cm]{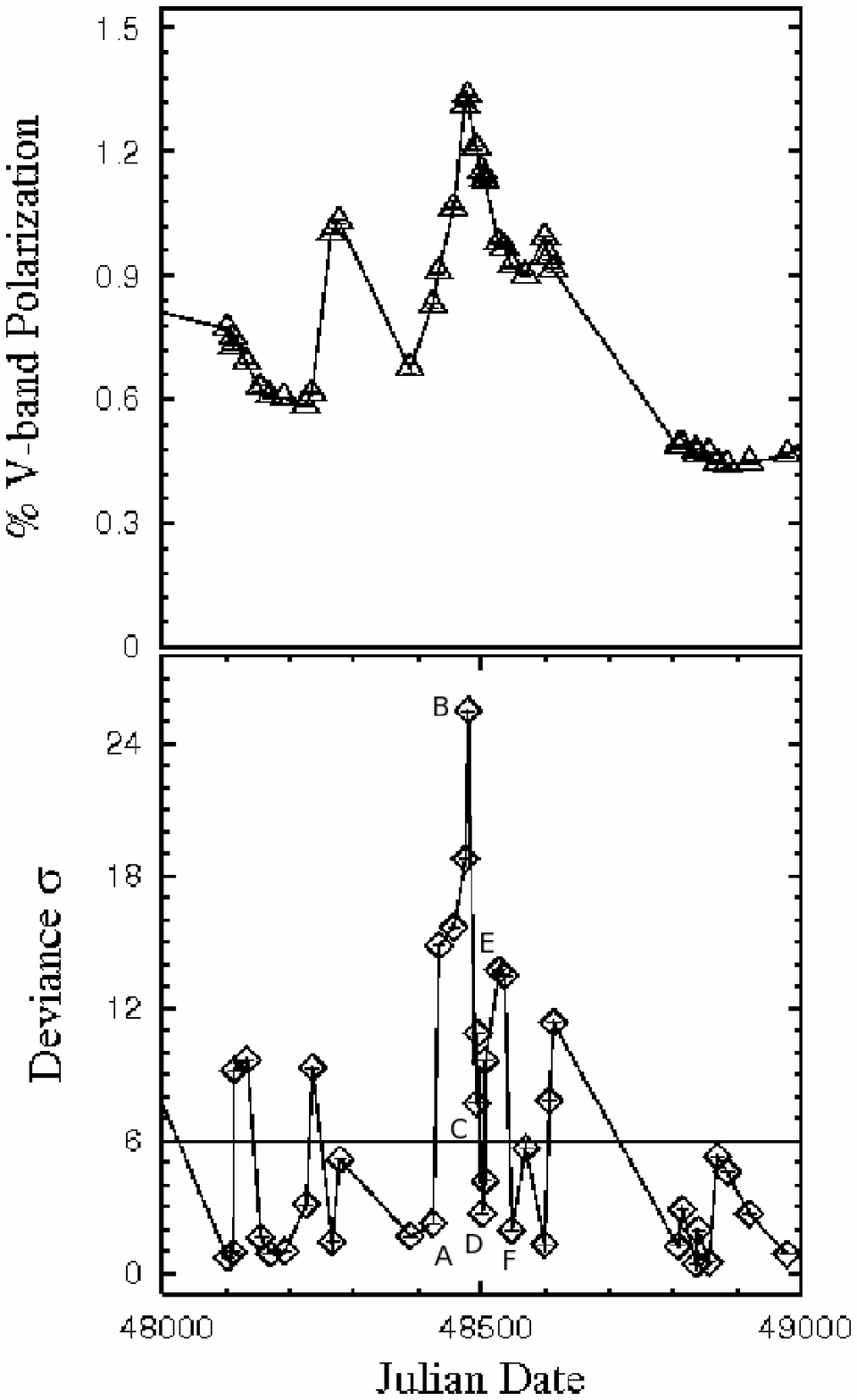}
\caption{The absolute value of the error weighted deviation (deviance) of every V-band polarization observation of $\pi$ Aqr from the best fit line in the star's Q-U diagram 
(Figure \ref{piaqrqu}) is plotted as a function of time (bottom panels) along with the V-band polarization as a function of 
time (top panel).  The right-side set of figures, a zoomed view of the entire (left-side figures) dataset, illustrates that the majority of the large 
deviations in the deviance correspond to one of the major polarimetric outbursts observed in $\pi$ Aqr.  The rise of the first component of the 
outburst (points A to B, $<$ 54 days) is similar to that of the second component of the outburst (points D to E, $<$ 35 days), while the time-scale of the decline 
phase in each component is also similar (points B to C, $<$ 13 days; points E to F, $<$ 10 days).  We interpret these data as evidence of departures from 
axisymmetry in the disk and postulate that they may be evident of newly injected disk material running into a density structure such as a spiral density 
wave (Figure \ref{funky}), or of the injection and subsequent circularization of new blobs into the inner disk, similar to the phenomenon noted by 
\citet{car07}.  \label{piaqrvar}}
\end{center}
\end{figure}

To further explore the origin of these PA deviations, we examine the deviance values computed for Figure \ref{var} as a function of time and 
compare the trends in these data to the temporal evolution of the V-band polarization.  Several clear trends are seen in the data for $\pi$ 
Aqr (Figure \ref{piaqrvar}): 1) a substantial number of deviance values exceeding the 6$\sigma$ level coincided with the latter-portion of 
the $\sim$390 day polarimetric outburst noted in Section \ref{piaqrpol}; and 2) there was no corresponding increase in the 
PA deviance coinciding with the $\sim$730 day polarimetric outburst reported in Section \ref{piaqrpol}.  Inspection of the right panel of Figure 
\ref{piaqrvar}, which depicts the behavior of the PA deviance during the $\sim$390 day outburst, reveals tentative evidence of two separate components 
in the deviance.  The rise of the first component of the deviance outburst (points A to B; Figure \ref{piaqrvar}) spanned $<$ 54 days while 
the decline of this first component (points B to C; Figure \ref{piaqrvar}) spanned $<$ 13 days.  The time-scale of the second component 
exhibited similarities, with a rise time (points D to E; Figure \ref{piaqrvar}) of $<$ 35 days and a decline time (points E to F; 
Figure \ref{piaqrvar}) of $<$ 10 days.  We note that the frequency of our observations do not trace the decline phase at a sufficient sampling rate 
to diagnose the precise time-scale of the decline; rather, the time-scales we quote are merely upper limits.  We can not exclude the possibility that the PA 
deviations returned to normal faster than the 10-13 days we quote.

\begin{figure}
\begin{center}
\includegraphics[width=8cm]{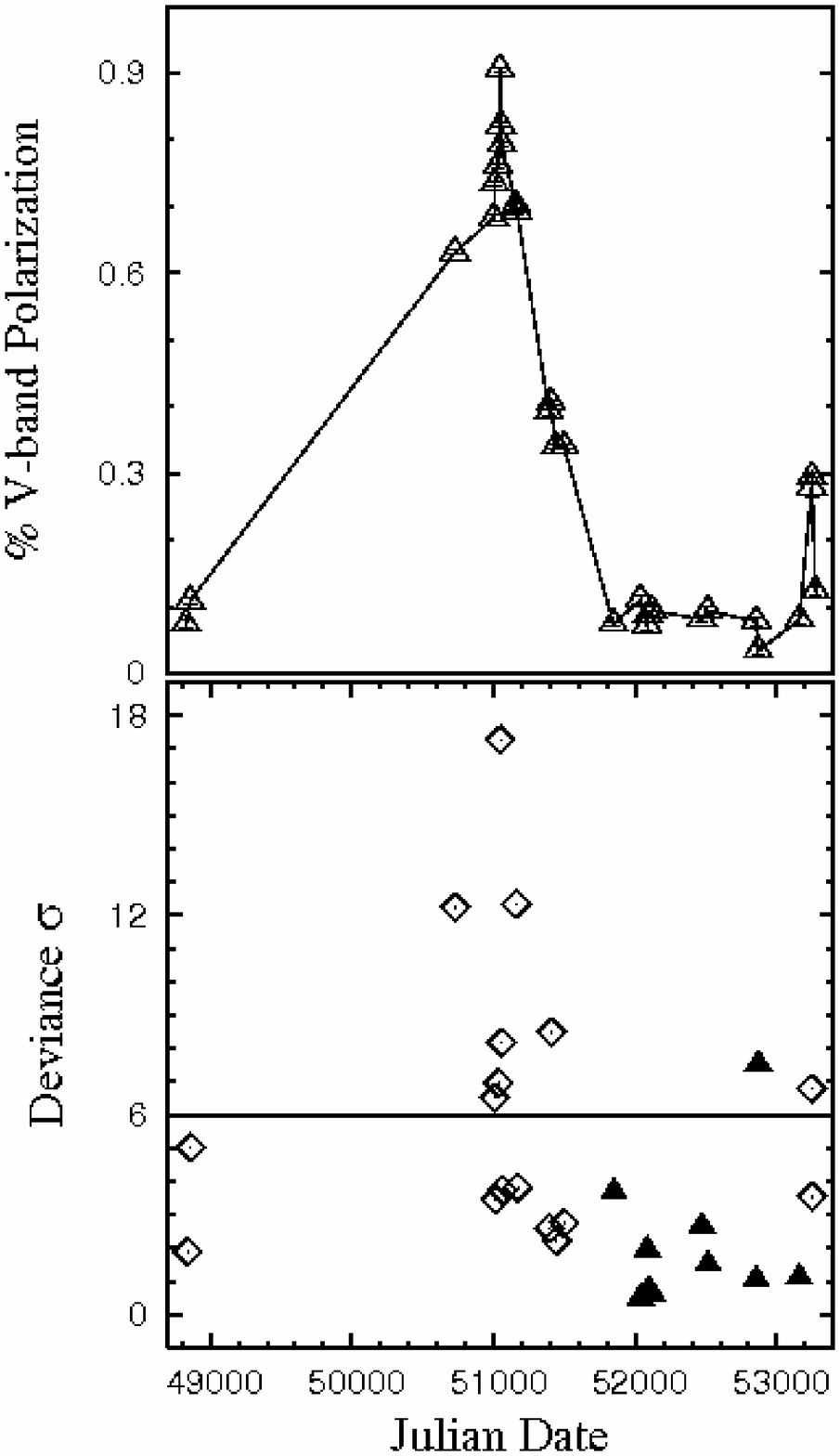}
\includegraphics[width=8cm]{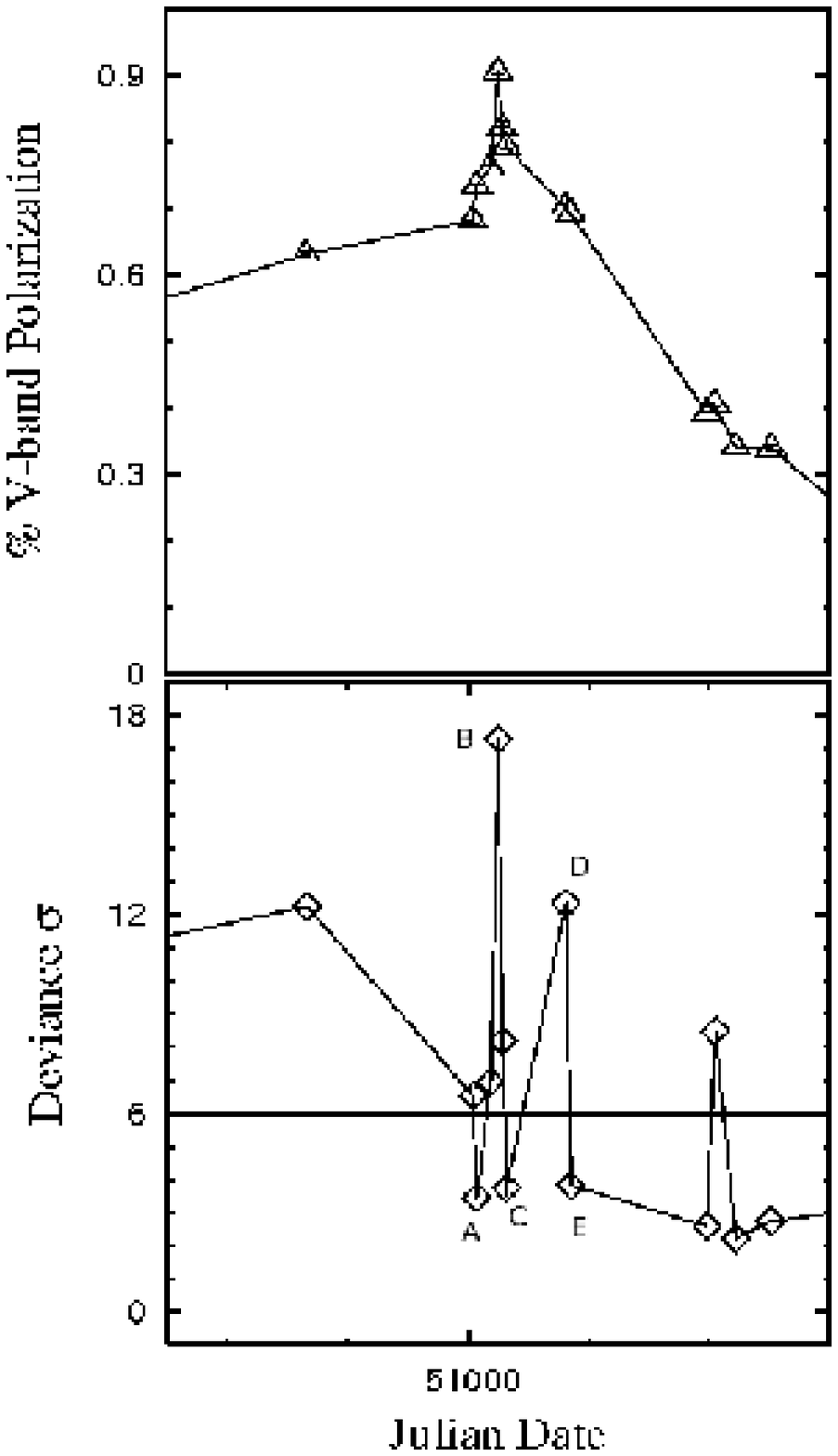}
\caption{The absolute value of the error weighted deviation (deviance) of every V-band polarization observation of 60 Cyg from the best fit line in the star's Q-U diagram 
(Figure \ref{60cygqu}) is plotted as a function of time (bottom panel) along with the V-band polarization as a function of 
time (top panel).  The right-side set of figures, a zoomed view of the entire (left-side figures) dataset, illustrates that the majority of the large 
deviations in the deviance correspond to large polarimetric outburst observed in 60 Cyg.  Both the time-scale of the rise in the outburst 
(points A to B, $<$ 26 days) and the time-scale of the decline in the outburst (points B to C, $<$ 14 days) are similar to the time-scales in the analogous phases 
observed for $\pi$ Aqr.  We interpret these data as evidence of departures from 
axisymmetry in the disk and postulate that they may evidence of newly injected disk material running into a density structure such as a spiral density 
wave (Figure \ref{funky}) or of the injection and subsequent circularization of new blobs into the inner disk, similar to the phenomenon noted by 
\citet{car07}.  \label{60cygvar}}
\end{center}
\end{figure}

A similar trend is observed in the temporal evolution of the PA deviance of the 60 Cyg data (see Figure \ref{60cygvar}).  A large outburst is 
seen in the deviance which coincides with the epoch of the single polarimetric outburst noted in Section \ref{60cygpol}.  Although the 
sampling of our 60 Cyg polarimetry is much more coarse compared to $\pi$ Aqr, Figure \ref{60cygvar} illustrates that the 
$<$ 26-day rise of the deviance outburst (points A to B) and subsequent $<$ 14-day decline (points B to C) qualitatively matches the rise and decline 
time-scales observed for $\pi$ Aqr.

\subsubsection{PA Changes: Evidence of Structure in the Inner Disk or Blob Injections?}

Variability in the intrinsic polarization position angles of classical Be stars is not typically expected for a standard axisymmetric disk 
system, although short-term \citep{car07} and long-term \citep{car07,tan07} changes have been reported.  \citet{car07} attributed variability on 
time-scales of hours to weeks in the Achernar system as evidence of the injection of blobs into the inner disk, leading to changes in the 
observed level of polarization and small changes in the polarization position angle which persist on order several weeks until the blob is 
circularized into a ring.  By contrast, \citet{tan07} attribute the decade time-scale evolution of Pleione's intrinsic polarization PA 
to disk precession caused by the system's binary companion.  We interpret the short-term variability (tens of days) of the intrinsic polarization position angle of both $\pi$ Aqr and 60 Cyg as evidence of 
low-amplitude, temporary departures from a simple axisymmetric disk density within the inner disk regions (several stellar radii) of these 
systems.

The deviations from axisymmetry we observe could be diagnostics of the injection of new blobs into the inner disk region, which 
subsequently circularize into a ring, thereby acting as an effective warp of the inner disk (see e.g. Figure \ref{funky}).  \citet{car07} reported variability in the polarization magnitude and position angle of the Be star Archernar, and offered a similar interpretation.  These authors noted that the circularization of a single blob would occur on time-scales of days and that the time-scale for a ring to incorporate into the inner disk region 
would be several weeks.  The newly injected material would then be re-accreted onto the central star if the mass-loss episode was a singular event,
whereas a steady stream of mass-loss episodes would cause the disk size and density to grow over time \citep{car07}.  
As the PA variations in the events observed in both $\pi$ Aqr and 60 Cyg grew over time-scales of 1-2 months and decayed in $\le$ 1-2 weeks, we speculate that perhaps the departure from axi-symmetry we witnessed was caused by the injection of multiple blobs into these systems which circularized
into a ring and were later incorporated into the inner disk region.  

Alternatively, the observed deviations from axisymmetry could be caused by a blob or series of blobs launched from a non-equatorial latitude from 
the stellar photosphere (Figure \ref{funky}).  Such blob(s) would still circularize on time-scales of days to weeks, but owing to the latitude of their ejection they would circularize into an inclined (non-coplanar) orbit with respect to the sem-major axis of the bulk of the disk material.  One could imagine this inclined ring eventually incorporating itself into the plane of the bulk of the disk material owing to viscous drag (or other) forces.  Thus we speculate that perhaps our observation of a temporary deviation in the polarization position angle might be diagnosing the presence of a newly injected blob, which is subsequently circularized, at an inclined orbit with respect to the plane of the pre-existing disk material.  We plan to explore the feasibility of both of the above scenarios using models similar to those presented in \citet{car07} in a future publication.

We note that one of the polarimetric outbursts of $/pi$ Aqr, occuring in 1994 June, was not accompanied by a corresponding variation in the 
polarization position angle (see e.g. Figure \ref{piaqrvar}).  Unlike the other outbursts, our temporal coverage of the 1994 event was especially 
sparse.  $\pi$ Aqr's V-band polarization was elevated in observations obtained on 5 June 1994, 7 June 1994, and 22 June 1994.  However, the onset 
of this outburst is unkown, as no observations were obtained between 10 September 1993 and the elevated 5 June 1994 observation.  Similary, there was a $\sim$6 week gap between the last elevated observation on 22 June 1994 and the subsequent observation on 3 August 1994, which merely places an upper limit on the time-scale of the polarization decline (of $<$ 6 weeks).  Given this sparse data sampling, we speculate that it is possible that 
we  simply ``missed''  observing the injection events which would have produced variations in the PA, and instead only sampled the aftermath 
of these events in which newly injected material had already circularized in the disk.  Alternatively, it is also possible that not all mass injection 
events yield deviations from axi-symmetry.  Clearly, additional observations of more polarimetric outburst events, at a higher sampling frequency, are 
needed to better constrain this phenomenon.

\subsection{Long-term Disk Loss/Disk Renewal Behavior of 60 Cyg} \label{60cyglong}

While our long-term Lyot and HPOL polarimetric monitoring of $\pi$ Aqr only covered 1 complete cycle of a Be to normal-B transition, our polarimetric 
observations of 60 Cyg actually span multiple disk loss/disk renewal transitions.  Specifically, there is clear evidence from the long-term behavior of 
both the V-band polarization and H-$\alpha$ EW (see Figure \ref{60cygew}; also \citealt{kou00}) that the circumstellar disk diagnosed by our 1987 PBO Lyot polarimetric observation had dissipated by the early 1990s and that our polarimetric data from 1997-2000 was probing a physically new circumstellar disk.  Interestingly, after subtracting the ISP from both the PBO Lyot and PBO HPOL polarimetry, we find the exact same (intrinsic) polarization position angle in both data sets.  This indicates that the older (1ate 1980s) and most recently generated (1997-2000) disks shared the same equatorial plane on the sky, and suggests a similar formation mechanism was responsible for both events.  We note that this scenario is fundamentally different than the misaligned double-disk structure reported for Pleione by \citet{tan07}, whereby the authors suggested that the PA of the disk was 
precessing due to interactions with the star's binary companion.

\section{Future Work} \label{future}

We have shown that the relatively rare occurence of a disk-loss episode in classical Be stars provides a unique opportunity to constrain the 
interstellar polarization component along the line of sight to these stars in exquisite detail.  While we have analyzed the behavior of the total polarization of our dataset, our accurate ISP determination will facilitate the improved removal of the interstellar component and thereby enable the time evolution of the intrinsic polarization component of these data to be studied in detail.  We are currently analyzing the behavior of the intrinsic polarization of these stars and plan to present results of detailed modeling of these systems, using the Monte Carlo code of \citet{car06}, in a 
future publication.

We also encourage observers to follow-up on the results presented in Section \ref{pachanges}, which indicate that the polarization PA of 
classical Be star disks may exhibit small-scale deviations during polarization outbursts.  We have suggested that the origin of these 
deviations might be evidence of newly injected blobs in the disk which have not yet circularized into a ring, or evidence of new blobs which were 
launched from a non-equatorial latitude thereby projecting them into an inclined circular orbit with respect to the plane of the pre-existing disk 
matrial.  We encourage high time cadence observations which diagnose 
the inner regions of such disk systems, such as polarimetry and/or infrared spectroscopy, to further constrain the origin of this phenomenon.

\section{Summary}

\begin{figure}
\begin{center}
\includegraphics[width=16cm]{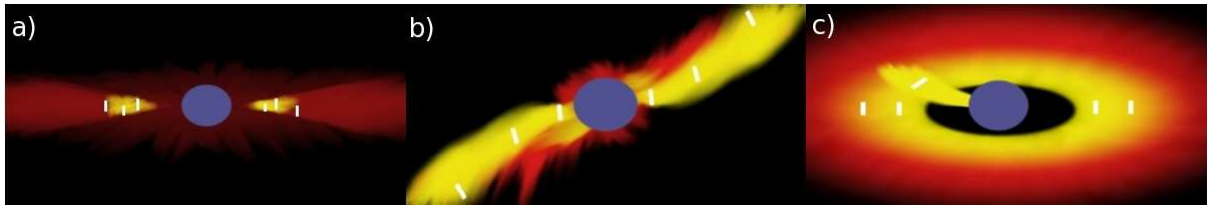}
\caption{\textbf{a)} A simple cartoon image depicting an edge-on Be star disk, whereby the inner region of the disk responsible for producing the observed 
intrinsic polarization is shaded yellow and the bulk of the region responsible for producing the observed H$\alpha$ emission is shaded red.  In the 
standard case of an axisymmetric disk, the maximum polarization vectors are oriented 90$^{\circ}$ from the plane of the geometrically 
thin disk.  \textbf{b)} Departures from axisymmetry in the inner-region of the disk, such as those caused by the inject of one or more blobs which circularize into rings, would lead to intrinsic polarization vectors which are not all aligned perpendicularily with the bulk PA of the disk.  We 
therefore speculate that one possible cause of the temporary (1-2 month) deviations in the intrinsic polarization PA values of both $\pi$ Aqr and 
60 Cyg is that blobs newly injected into the disk are created a warp-like structure in the inner disk. \textbf{c)} Alternatively, the observed deviations from axisymmetry could be caused by a blob (or series of blobs) launched from a non-equatorial latitude from the stellar photosphere, leading to the formation of a ring in an inclined (non-coplanar) orbit with respect to the semi-major axis of pre-existing disk material.\label{funky}}
\end{center}
\end{figure}

We have presented the results of 127 spectropolarimetric observations of the Be star $\pi$ Aqr spanning 15 years, and 35 spectropolarimetric 
and 65 H$\alpha$ spectroscopic observations of the Be star 60 Cyg spanning 14 years.  These data trace the evolution of each star as it transitions 
from having a gaseous circumstellar disk (``Be phase'') to a diskless state (``normal B-star phase'').  We find:

\begin{itemize}
\item  60 Cyg's disk emission strength, as measured by its H$\alpha$ EW, declined from maximum emission to pure absorption monotonically in $\sim$870 days.  The maximum observed H$\alpha$ EW lagged the maximum observed V-band polarization by $\sim$120 days, indicating that the total disk dissipation time-scale was $\sim$1000 days ($\sim$870 + $\sim$120),  and that the disk clearing proceeded in an ``inside-out'' fashion.  We found the dissipation of disk, as measured from the time-scale of the decrease in the observed polarization, occured over a viscous time-scale if we adopted a viscosity parameter $\alpha$ of 0.14.
\item $\pi$ Aqr's disk emission strength as measured by its H$\alpha$ EW declined from maximum emission to pure absorption over a time-frame of 
$\sim$2440 days.  Two extended outburst events lasting $\sim$390 and $\sim$730 days respectively were observed during $\pi$ Aqr's disk-loss phase, which had the 
effect of temporarily stalling the loss of the disk.  An observed lag in the onset of the minimum H$\alpha$ EW strength from the minimum V-band polarization 
indicates that $\pi$ Aqr's disk clearing also proceeded in an ``inside-out'' fashion.
\item The time-scale of the disk-loss events in 60 Cyg and $\pi$ Aqr corresponds to almost 6 and 29 complete orbits of each star's binary companion 
respectively.  This suggests that each star's binary companion does not influence the primary star (or its disk) in a manner similar to the highly 
eccentric $\delta$ Sco system \citep{mir01}. 
\item We fit the wavelength dependence of multi-epoch spectropolarimetric observations obtained during the quiescent diskless phase of each 
star with a modified Serkowski law to derive the interstellar polarization along the line of sight to each star.  We determine the best fit modified 
Serkowski law parameters for 60 Cyg to be P$_{max}$ = 0.112\% $\pm$ 0.001\%, $\lambda_{max}$ = 6977\AA, PA = 41.5$^{\circ}$, and K = 1.17, while for $\pi$ Aqr 
we derive the parameters P$_{max}$ = 0.514\% $\pm$ 0.001\%, $\lambda_{max}$ = 4959\AA, PA = 108.7$^{\circ}$, and K = 0.83.  
\item We find that the position angle of intrinsic polarization arising from 60 Cyg's disk is $\theta_{\ast}$ = 107.7$\pm$0.4$^{\circ}$, indicating that the disk is oriented on the sky at a position angle of $\theta_{disk}$ = 17.7$^{\circ}$ (measured North to East).  We find that the position angle of intrinsic polarization arising from $\pi$ Aqr's disk is $\theta_{\ast}$ = 166.7$\pm$0.1$^{\circ}$, indicating that the disk is oriented on the sky at a position angle of $\theta_{disk}$ = 76.7$^{\circ}$.
\item We detect clear evidence of small outburst events in the V-band polarization and H$\alpha$ EW during the quiescent diskless 
phase of both $\pi$ Aqr and 60 Cyg, which persist for several months in the V-band polarization.  As these events have similar time-scales as the blob injection events likely induced by non-radial pulsations in the Be star $\mu$ Cen \citep{han93,riv98}, we suggest these outburst events indicate the injection of one (or several) blobs of material from the stellar photosphere which are subsequently accreted back onto the central star and/or decreted on a viscous time-scale.
\item We also detect evidence of deviations from the intrinsic polarization position angle in both 60 Cyg and $\pi$ Aqr which persist for 1-2 months and coincide with large outbursts in the total V-band polarization, and confirm these events are real via an examination of the false discovery rate for each object.  For $\pi$ Aqr, we determine the magnitude of one of these deviations to be $\sim$2.43$^{\circ}$ from the bulk intrinsic polarization PA of the disk (166.7$^{\circ}$).  These results are indicative of deviations from a simple axisymmetric disk density structure in each system.  We 
propose that either we are witnessing the injection and subsequent circularization of new blobs into the inner disk region in a similar 
manner as noted by \citet{car07} or the injection and subsequent circularization of new blobs at an inclined orbit to the plane of pre-existing disk material (see Figure \ref{funky}).
\end{itemize}

\acknowledgments

We thank our referee, David Harrington, for providing helpful suggestions which improved the quality of this paper.  We also thank Brian Babler for his invaluable assistance with various aspects of PBO HPOL data, the PBO and Ritter science teams for providing observing and data reduction support, and Kenneth H. Nordsieck for providing access to the PBO Lyot and HPOL spectropolarimeters.  JPW thanks A. Carciofi and A Okazaki for providing helpful 
comments regarding the effects of binary companions in Be star systems.  JPW acknowledges support from NSF Astronomy \& Astrophysics Postdoctoral Fellowship AST 08-02230, and ZHD acknowledges partial support from the University of Washington Pre-MAP program.  HPOL observations were supported under NASA contract NAS5-26777 with the University of Wisconsin-Madison.  
Observations at the Ritter Observatory are supported by the NSF under the PREST grant AST 04-40784.  
This study has made use of the SIMBAD database, operated at CDS, Strasbourg, France, and the NASA ADS
service.

\newpage
\clearpage
\begin{table}
\begin{center}
\footnotesize
\caption{Summary of 60 Cyg Data \label{60cygdata}}
\begin{tabular}{lcccccc}
\tableline
Date & JD & Obs & H$\alpha$ EW & V-band Pol & PA & State \\
\nodata & \nodata & \nodata & \AA\ & \% & degrees & \nodata \\
\tableline
1992 Aug 3 & 2448837 & PBO &  4.5 $\pm$ 0.1 & 0.074 $\pm$ 0.006 & 58.2 & H \\
1992 Aug 31 & 2448865 & PBO &  3.5 $\pm$ 0.1 & 0.108 $\pm$ 0.004 & 52.2 & H \\
1997 Nov 10 & 2450732 & PBO & -5.5 $\pm$ 0.1 & 0.631 $\pm$ 0.003 & 102.1 & H  \\
\tableline
\end{tabular}
\end{center}
\vspace{-0.3in}
\tablecomments{A sample summary of spectroscopic and spectropolarimetric observations of 60 Cygni; the full version of this Table 
is available in the online-edition of this Journal.  The third column denotes the location at which the data were obtained (PBO = Pine Bluff 
Observatory, RIT = Ritter Observatory), while the seventh column denotes a qualitative assessment of whether 60 Cyg is in a polarimetrically
high state (H), indicative of the presence of a strong disk, or in a polarimetrically low state (L), in which evidence of the presence of a disk is 
absent (see Section \ref{serk} for further discussion of these designations).  Note that 
we have labeled the two observations in 1992 August as being ``high state'', primarily because they are isolated from subsequent observations by 
$\sim$5 years.}
\end{table}

\newpage
\clearpage
\begin{table}
\begin{center}
\footnotesize
\caption{Summary of $\pi$ Aqr Data \label{piaqrdata}}
\begin{tabular}{lcccccc}
\tableline
Date & JD & Obs & H$\alpha$ EW & V-band Pol & PA & State \\
\nodata & \nodata & \nodata & \AA\ & \% & degrees & \nodata \\
\tableline
1985 May 15 & 2446200 & McD$^{1}$ & \nodata & 1.53$\pm$0.03 & 158.8 & \nodata \\
1986 Jun 15 & 2446596 & McD$^{2}$ & \nodata & 1.47$\pm$0.02 & 159.1 & \nodata \\
1987 Aug 15 & 2447022 & McD$^{2}$ & \nodata & 1.50$\pm$0.03 & 157.4 & \nodata \\
1988 Jul 15 & 2447357 & McD$^{2}$ & \nodata & 1.04$\pm$0.06 & 155.0 & \nodata \\
1989 Jun 15 & 2447692 & McD$^{2}$ & \nodata & 0.97$\pm$0.04 & 152.1 & \nodata \\
1989 Aug 8 & 2447747 & PBO & -23.5$\pm$0.1  & 1.277$\pm$0.005 & 156.5 & H \\
1989 Oct 26 & 2447826 & PBO & -24.6$\pm$0.1 & 0.879$\pm$0.003 & 148.9 & H \\
\tableline
\end{tabular}
\end{center}
\vspace{-0.3in}
\tablecomments{A sample summary of spectroscopic and spectropolarimetric observations of $\pi$ Aquarii; the full version of this Table 
is available in the online-edition of this Journal.  The third column denotes the location at which the data were obtained (PBO = Pine Bluff 
Observatory, McD = McDonald Observatory), while the seventh column denotes a qualitative assessment 
of whether $\pi$ Aqr is in a polarimetrically high state (H), indicative of the presence of a strong disk, or in a polarimetrically low state (L), in which evidence of the presence of a disk is absent (see Section \ref{serk} for further discussion of these designations).  $^{1}$ data are from \citet{mcd86}, $^{2}$ data are from \citet{mcd90}.}
\end{table}

\newpage
\clearpage
\begin{table}
\begin{center}
\footnotesize
\caption{PBO Lyot Polarimetry \label{lyot}}
\begin{tabular}{lccccc}
\tableline
Target Star & JD & Date & \%Pol. & P.A. & Wavelength Coverage \\
\nodata & \nodata & \nodata & \% & degrees & \AA\ \\
\tableline
60 Cyg & 2446998 &  1987 Jul 22 & 0.44 $\pm$ 0.02 & 104 & 4600-7200 \\
$\pi$ Aqr & 2444116 &  1979 Aug 31 & 1.17 $\pm$ 0.03 & 153 & 4800-7600  \\
$\pi$ Aqr & 2444131 &  1979 Sep 15 & 1.13 $\pm$ 0.02 & 155 & 4800-7600  \\
$\pi$ Aqr & 2444133 &  1979 Sep 17 & 1.16 $\pm$ 0.03 & 156 & 4800-7600 \\
$\pi$ Aqr & 2444138 &  1979 Sept 22 & 1.12 $\pm$ 0.02 & 157 & 4800-7600  \\
$\pi$ Aqr & 2444439 &  1980 Jul 19 & 1.31 $\pm$ 0.12 & 159 & 4800-8200  \\
\tableline
\end{tabular}
\end{center}
\vspace{-0.3in}
\tablecomments{Summary of polarization measurements of 60 Cyg and $\pi$ Aqr obtained between 1979-1987 using the Lyot Polarimeter at the 
University of Wisconsin's Pine Bluff Observatory 36'' telescope.}
\end{table}

\newpage
\clearpage
\begin{table}
\begin{center}
\footnotesize
\caption{ Field Star Polarization Data \label{t2}}
\begin{tabular}{lccccc}
\tableline
Target Star & Field Star & Spectral Type & Distance(pc) & \%Pol. & P.A. \\
\tableline
$\pi$ Aqr & HD213789 &  G6III & 136 & 0.53$^{1}$ & 133.7$^{1}$ \\
$\pi$ Aqr & HD213119 &  K5III & 178 & 0.23$^{1}$ & 110.4$^{1}$ \\
$\pi$ Aqr & HD212320 &  G6III & 141 & 0.44$^{1}$ & 160.5$^{1}$ \\
$\pi$ Aqr & HD211924 &  B5IV & 282 & 1.16$^{1}$ & 140.2$^{1}$ \\
$\pi$ Aqr & HD211838 &  B8III & 228 & 0.46$^{1}$ & 138.8$^{1}$ \\
$\pi$ Aqr & HD211304 &  B9 & 306 & 0.22$^{1}$ & 107.6$^{1}$ \\
$\pi$ Aqr & HD211099 &  B9 & 455 & 0.26$^{1}$ & 119.2$^{1}$ \\
60 Cyg & HD198915 & B6V & 288 & 0.19$^{2}$ & 15.0$^{2}$ \\
\tableline
\end{tabular}
\end{center}
\vspace{-0.3in}
\tablecomments{Note that spectral types were obtained from SIMBAD, and distance measurements were obtained from the 
Hipparcos catalog \citep{per97}.  For reference, the Hipparcos derived distances to $\pi$ Aqr and 60 Cyg are 340pc and 418pc respectively.  
Archival polarization data were compiled from $^{1}$\citet{mat70} and $^{2}$\citet{hei00}.}
\end{table}

 \newpage
\clearpage
\begin{table}
\begin{center}
\footnotesize
\caption{Summary of Interstellar Polarization Values \label{isptable}}
\begin{tabular}{lcccccc}
\tableline
Star & Value & P$_{max}$ & $\lambda_{max}$ & PA & K & C \\
\tableline

60 Cyg & Q$^{'}$ & -0.076 $\pm$ 0.000 & 6977 $\pm$ 241 & 135.0 & 1.17 & 0 \\
60 Cyg & U$^{'}$ & -0.083 $\pm$ 0.001 & 6977 $\pm$ 241 & 90.0 & 1.17 & 0 \\
60 Cyg & P & 0.112  $\pm$ 0.001 & 6977 $\pm$ 241 & 41.5 & 1.17 & 0 \\
Pi Aqr & Q$^{'}$ & -0.265 $\pm$ 0.001 & 4959 $\pm$ 15 & 135.0 & 0.83 & 0.04 \\
Pi Aqr & U$^{'}$ & -0.462 $\pm$ 0.001 & 4959 $\pm$ 15 & 90.0 & 0.83 & 0 \\
Pi Aqr & P & 0.514 $\pm$ 0.001 & 4959 $\pm$ 15 & 108.7 & 0.83 & 0 \\
\tableline
\end{tabular}
\end{center}
\vspace{-0.3in}
\tablecomments{The derived modified Serkowski law parameters \citep{ser75,wil82} derived for the perpendicular (U$^{'}$) and 
parallel (Q$^{'}$) ISP components.  The total ISP estimate for each star was computed by vectoral addition of the quoted parallel and 
perpendicular components.  Note that column 7, ``C'', denotes a constant term added to the modified Serkowski-law fit of the 
$\pi$ Aqr parallel ISP component to accomodate the presence of a possible small intrinsic polarization component.}
\end{table}

\newpage
\clearpage
 
\newpage
\clearpage

\newpage
\clearpage

\newpage
\clearpage
 
\newpage
\clearpage

\newpage
\clearpage

\newpage
\clearpage

\newpage
\clearpage

\newpage
\clearpage

\newpage
\clearpage
 
\newpage
\clearpage

\newpage
\clearpage

\newpage
\clearpage

\newpage
\clearpage
 
\end{document}